\documentclass[lettersize,journal]{IEEEtran}
\usepackage{amsmath,amsfonts}
\usepackage{algorithmic}
\usepackage{algorithm}
\usepackage{array}
\usepackage[caption=false,font=normalsize,labelfont=sf,textfont=sf]{subfig}
\usepackage{textcomp}
\usepackage{stfloats}
\usepackage{url}
\usepackage{verbatim}
\usepackage{graphicx}
\usepackage{cite}
\usepackage{multirow}
\usepackage{booktabs}
\usepackage{siunitx}
\usepackage{hyperref}

\usepackage{xcolor}

\begin{document}

\title{Dynamic Convolutional Neural Networks \\ as Efficient Pre-trained Audio Models}

\author{Florian Schmid$^1$, Khaled Koutini$^2$, and Gerhard Widmer$^{1,2}$ \\
        % <-this % stops a space
$^1$\textit{Institute of Computational Perception}, $^2$\textit{LIT Artificial Intelligence Lab}  \\
\textit{Johannes Kepler University}, Linz, Austria \\
\{florian.schmid, khaled.koutini\}@jku.at
%\thanks{More on the authors...}
%\thanks{DOI...}
\thanks{Code available: \href{https://github.com/fschmid56/EfficientAT}{https://github.com/fschmid56/EfficientAT}}
}
% <-this % stops a space

\markboth{IEEE/ACM TRANSACTIONS ON AUDIO, SPEECH, AND LANGUAGE PROCESSING}%
{Shell \MakeLowercase{\textit{et al.}}: A Sample Article Using IEEEtran.cls for IEEE Journals}


\maketitle

\begin{abstract}
The introduction of large-scale audio datasets, such as AudioSet, paved the way for Transformers to conquer the audio domain and replace CNNs as the state-of-the-art neural network architecture for many tasks. Audio Spectrogram Transformers are excellent at exploiting large datasets, creating powerful pre-trained models that surpass CNNs when fine-tuned on downstream tasks. However, current popular Audio Spectrogram Transformers are demanding in terms of computational complexity compared to CNNs. Recently, we have shown that, by employing Transformer-to-CNN Knowledge Distillation, efficient CNNs can catch up with and even outperform Transformers on large datasets. In this work, we extend this line of research and increase the capacity of efficient CNNs by introducing dynamic CNN blocks, constructed of dynamic non-linearities, dynamic convolutions and attention mechanisms. We show that these dynamic CNNs outperform traditional efficient CNNs, in terms of the performance--complexity trade-off and parameter efficiency, at the task of audio tagging on the large-scale AudioSet. Our experiments further indicate that the introduced dynamic CNNs achieve better performance on downstream tasks and scale up well, attaining Transformer performance and even outperforming them on AudioSet and several downstream tasks.  

\end{abstract}


\section{Introduction}

Pre-trained deep neural networks have emerged as a pivotal paradigm in the field of machine learning over the last few years. Leveraging transfer learning techniques, pre-trained models can significantly enhance the performance on downstream tasks, in particular, when the training data is insufficient to learn an end-to-end model from scratch. Such models are typically pre-trained in a supervised or self-supervised fashion on large datasets, such as ImageNet~\cite{Deng09ImageNet} in the vision domain, or AudioSet~\cite{audioset2017Gemmeke} in the audio domain. While convolutional neural networks (CNNs) have been the method of choice during the past years to create pre-trained models in both the audio and the vision domain, Transformers~\cite{Vaswani17Attention} recently surpassed them due to their ability to scale up and exploit large datasets~\cite{Dosovitskiy20Image, Gong21Ast}. With superior performance on the large pre-training datasets, Transformers then overtook CNNs also on many downstream tasks with smaller datasets. However, Transformers are computationally costly in training and inference, as the attention operation scales quadratically with respect to the processed sequence length. For this reason, CNNs maintain their dominance on resource-constrained platforms, such as mobile devices.   

Recently, it has been shown in the audio domain that efficient CNNs can attain and even outperform Transformers on large-scale datasets when they are trained using Knowledge Distillation (KD)~\cite{Hinton2015distilling, Bucila06Compression, Ba14KD} from a Transformer ensemble. Concretely, Schmid et al.~\cite{Schmid22Efficient} train MobileNetV3s (MNs)~\cite{Howard19MobileNetV3} on AudioSet
% ~\cite{audioset2017Gemmeke} GW: has been cited above
using offline KD from an ensemble consisting of 9 different Patchout FaSt Spectrogram Transformer (PaSST)~\cite{Koutini21Passt} models. The resulting efficient pre-trained MNs have been shown to extract high-quality general-purpose audio embeddings that can generalize to downstream tasks in various audio domains such as music, speech and environmental sounds~\cite{schmid2023low}. Compared to Transformers, the quality of  extracted audio embeddings is comparable while the computational cost of inference is much lower. Although the MNs achieve excellent performance on AudioSet~\cite{Schmid22Efficient} and serve as high-quality audio embeddings extractors~\cite{schmid2023low}, they remain to be tested in end-to-end fine-tuning, in which a pre-trained model is directly fine-tuned on a downstream task.

In this work, we extend this line of research and propose computationally efficient pre-trained audio model, obtained by integrating dynamic components into MobileNetV3s (DyMNs). We use the Knowledge Distillation pre-training setup of \cite{Schmid22Efficient} and show that the proposed DyMNs achieve substantially higher pre-training performance on AudioSet compared to MNs. We train MNs and DyMNs of different complexity levels on the downstream tasks of polyphonic musical instrument recognition \textit{(OpenMic dataset}~\cite{humphrey2018openmic}), environmental sound classification \textit{(ESC50 dataset}~\cite{piczak2015esc}), acoustic scene classification \textit{(TAU Urban Acoustic Scenes 2020 challenge}~\cite{heittola2020acoustic}), and sound event tagging \textit{(FSD50K}~\cite{fonseca2021fsd50k}). The results show that MNs and DyMNs can attain or even surpass the performance of a single teacher model PaSST~\cite{Koutini21Passt} on many downstream tasks, while being much more computationally efficient. The proposed DyMNs outperform MNs for the majority of downstream tasks and complexity levels.

The motivation for integrating dynamic components into MNs is threefold. Firstly, instead of scaling CNNs by network width and depth, which increases the computational complexity substantially, a variety of lightweight dynamic components such as dynamic non-linearities~\cite{chen2020dynamic_relu, si2018dynamic}, convolutions~\cite{chen2020dynamic_conv, yang2019condconv} and attention mechanisms~\cite{Hu18Squeeze, hou2021coordinate, lee2019srm, woo2018cbam, park2018bam, misra2021triplet, cao2019gcnet} have been proposed. These dynamic components have been shown to increase the performance while only marginally adding to the computational complexity in terms of consumed multiply-accumulate operations (MACs) at inference time. Secondly, the success of Transformers is largely based on self-attention, a highly dynamic operation that adapts its attention weights based on input data. Thirdly, an ablation study conducted in \cite{Schmid22Efficient} revealed that Squeeze-and-Excitation~\cite{Hu18Squeeze}, a component that dynamically computes channel attention weights based on input data, is an integral part of MNs to achieve high performance on AudioSet~\cite{audioset2017Gemmeke}. 

Concretely, our proposed DyMN block deviates from the traditional residual inverted bottleneck block~\cite{Sandler18MobileNetsV2} in MNs in three ways: we include dynamic convolution~\cite{chen2020dynamic_conv}, replace the non-linearity after the depthwise convolution with dynamic ReLU~\cite{chen2020dynamic_relu}, and make use of Coordinate Attention~\cite{hou2021coordinate} instead of Squeeze-and-Excitation~\cite{Hu18Squeeze}. Since all these dynamic components perform operations based on the context of the input sample, we compute the context once and share it across all dynamic components in a block.

To summarize our contribution, we extend the work of Schmid et al.~\cite{Schmid22Efficient} and introduce a dynamic CNN block to further boost the pre-training performance of efficient CNNs on AudioSet~\cite{audioset2017Gemmeke}. We show that the proposed dynamic CNNs outperform traditional efficient CNNs on downstream tasks in an end-to-end fine-tuning setting. The proposed dynamic CNNs attain and even outperform current Audio Spectrogram Transformers on several downstream tasks while being more computationally efficient.

%To summarize our contribution, firstly, we extend the work of Schmid et al.~\cite{Schmid22Efficient} and show that CNNs trained with Transformer-to-CNN Knowledge Distillation can attain Transformer performance on various downstream tasks in the audio domain \fs{in an end-to-end fine-tuning setting}. Secondly, we introduce a dynamic CNN block to further boost the pre-training performance of efficient CNNs on AudioSet~\cite{audioset2017Gemmeke}. Thirdly, we show that the proposed dynamic CNNs outperform traditional CNNs on downstream tasks and perform favorably compared to current Audio Spectrogram Transformers on many tasks.

The remainder of this paper is structured as follows. Related work is reviewed in Section~\ref{sec:related}, followed by the introduction of the proposed dynamic model in Section~\ref{sec:dynamic}. The pre-training setup and results are presented in Section~\ref{sec:pre-training}; Section~\ref{sec:experiments} then shows the experiments and results on the downstream tasks. A detailed systematic configuration study and an analysis of the dynamic components are performed in Sections~\ref{sec:ablation} and \ref{sec:inspecting}, respectively. The paper is concluded in Section~\ref{sec:conlusion}. 

%\gw{\footnote{A note to the reviewers of this paper: to contain the length of this submission, the mentioned Section VII is only available in a longer pre-print version of this paper, available via arxiv. We will explain this to the reader in a final version of the paper.}}

\section{Related Work}
\label{sec:related}

In this section, the related work is covered. We start with a general recap on efficient CNN architectures and popular dynamic CNN components that were introduced to further increase a CNN's computational efficiency. We then cover the literature on pre-trained audio models to set the stage for introducing our dynamic CNN serving as an efficient, pre-trained audio model.

\subsection{Efficient CNN Architectures}

Much effort has been invested in research on designing efficient CNNs, such as the series of MobileNets~\cite{Howard17MobileNets, Sandler18MobileNetsV2, Howard19MobileNetV3}, EfficientNets~\cite{Tan19EfficientNet, Tan21EfficientNetV2} or ShuffleNets~\cite{zhang2018shufflenet, ma2018shufflenetv2}. MobileNetV1~\cite{Howard17MobileNets} substantially reduces the computational complexity of conventional convolution layers by factorizing them into a depthwise and a 1x1 pointwise convolution. On top of this, MobileNetV2~\cite{Sandler18MobileNetsV2} introduces inverted residual blocks with linear bottlenecks, leading to better accuracy and computational efficiency. MobileNetV3~\cite{Howard19MobileNetV3} additionally adds Squeeze-and-Excitation~\cite{Hu18Squeeze} layers after the depthwise filters, upgrades activation functions using swish non-linearity~\cite{ramachandran2017searching} and optimizes the global network structure using platform-aware network architecture search~\cite{tan2019mnasnet}. EfficientNet~\cite{Tan19EfficientNet} builds on the MobileNetV2 inverted residual block and introduces compound scaling laws of depth, width and input resolution. Similar to MobileNetV3, EfficientNetV2~\cite{Tan21EfficientNetV2} performs a neural architecture search to optimize parameter efficiency and training speed. Originally introduced in the vision domain, MobileNets and EfficientNets have been shown to provide a good performance--complexity trade-off also in the audio domain~\cite{Kong20PANNs, Gong21PSLA, Gong22CMKD, Schmid22Efficient}. 

\subsection{Dynamic CNN Components}

While scaling CNNs by width and depth typically improves performance, it significantly increases the model's complexity. In particular, the computational demand of a CNN scales with the square of the model's width. As an alternative strategy, a lot of research has focused on using a fixed number of channels more efficiently by introducing dynamic components to CNNs. While the majority of work proposes attention mechanisms~\cite{Hu18Squeeze, hou2021coordinate, lee2019srm, misra2021triplet, cao2019gcnet, park2018bam, woo2018cbam}, others have used dynamic convolutions~\cite{chen2020dynamic_conv, wu2019dynamic_conv_pay, verelst2020dynamic_conv_spatial, yang2019condconv, zhang2020dynet} and dynamic non-linearities~\cite{si2018dynamic, chen2020dynamic_relu}. 

\textbf{Attention Mechanisms:} The most prominent instance of a CNN attention mechanism is Squeeze-and-Excitation (SE)~\cite{Hu18Squeeze} being integrated into MobileNetV3~\cite{Howard19MobileNetV3} and EfficientNets~\cite{Tan19EfficientNet,Tan21EfficientNetV2}. As shown in Eq. \ref{eq:se}, SE applies a squeeze ($F_{\mathit{sq}}$) and an excitation ($F_{\mathit{ex}}$) operation on a feature map to obtain channel recalibration weights $\mathbf{s}$. $F_{\mathit{sq}}$ applies global average pooling (GAP) to collect contextual information while $F_{\mathit{ex}}$ captures channel-wise dependencies via a learnable non-linear transformation followed by a sigmoid activation to compute weights.

\begin{equation}
\label{eq:se}
    \begin{split}
    \mathbf{s} = F_{\mathit{ex}}(F_{\mathit{sq}}(\mathbf{x}), \textbf{W})
    \end{split}
\end{equation}

Other attention mechanisms differ mainly in how they realize $F_{\mathit{sq}}$ and $F_{\mathit{ex}}$. The Style-based Recalibration Module (SRM)~\cite{lee2019srm} extends SE by combing GAP and global standard pooling to realize $F_{\mathit{sq}}$, followed by a channel-wise fully connected layer, batch normalization and a sigmoid activation function for $F_{\mathit{ex}}$. CBAM~\cite{woo2018cbam} additionally computes attention weights for spatial locations using a channel pooling operation to squeeze contextual information. Coordinate Attention (CA)~\cite{hou2021coordinate} factorizes the attention mechanism into two separate context encoding processes of the spatial dimensions. Features are aggregated by GAP along the spatial dimensions, processed separately and then used as spatially aware recalibration weights. Triplet Attention (TA)~\cite{misra2021triplet} reduces the three dimensions of a feature map by average and max pooling, processes the three 2-dimensional slices separately, and recalibrates the feature map with the resulting three sets of recalibration weights. Global Context (GC) blocks~\cite{cao2019gcnet} differ from the aforementioned attention mechanisms by performing additive instead of multiplicative recalibration. The recently introduced Global Response Normalization (GRN)~\cite{woo2023convnextv2} serves as a particularly lightweight attention module by recalibrating the channels based on their L2-norms and adding only two learnable parameters, a scale and a shift.  

\textbf{Dynamic Convolution:} In contrast to standard convolution layers, dynamic convolutions adapt the kernel weights based on global context information extracted from an input. Early approaches in the vision domain have explored generating the kernels directly~\cite{jia2016dynamic, klein2015dynamic}, resulting in a substantial increase in complexity as the number of parameters of convolution kernels is large. A more lightweight approach is to predict coefficients to linearly combine a fixed set of kernels. In this spirit, CondConv~\cite{yang2019condconv} computes a linear mixture of $K$ distinct trainable kernel weights. As shown in Eq. \ref{eq:conv_mixture}, only a single convolution ($*$) with the mixed kernel has to be performed, as this is equivalent to combining the results of $K$ individual convolutions. The weights $\alpha_i$ per filter are computed dynamically from the input using GAP, a linear transformation, and a sigmoid activation.

\begin{equation}
\label{eq:conv_mixture}
    \begin{split}
    \alpha_1 (W_1 * x) + \alpha_2 (W_2 * x) + ... + \alpha_K (W_K * x) = \\ (\alpha_1 W_1 + \alpha_2 W_2 + ... + \alpha_K W_K) * x 
    \end{split}
\end{equation}

DyNet~\cite{zhang2020dynet} proposes a similar dynamic convolution mechanism, motivated from the perspective of extracting noise-irrelevant features, showing that dynamic filters generate more diverse feature maps with lower cross-correlation. A common global context is shared across CNN blocks, e.g., the global context extracted by GAP from a block input is used to parameterize the three dynamic convolutions in an inverted bottleneck block as used in MobileNetV2s~\cite{Sandler18MobileNetsV2}.
The dynamic convolution introduced in \cite{chen2020dynamic_conv} compresses the kernel space by introducing the constraint $\sum_{k}\alpha_k = 1$ on the computed kernel attention weights. As a result, a smaller number of kernels $K$ can be used, saving computations and parameters. The constraint is enforced by applying a softmax instead of a sigmoid activation. Temperature scaling is used before the softmax to ensure near-uniform attention in early epochs. 

Besides dynamic convolution in the vision domain, some works have explored generating input-aware convolutional filters based on the input sentences in NLP~\cite{wu2019dynamic_conv_pay, shen2017learning, gong2018convolutional}.  In the audio domain, temporal dynamic convolutions (TDY)~\cite{kim2022temporal} and frequency dynamic convolutions (FDY)~\cite{nam2022frequency} have gained popularity recently. TDY dynamically adapts the filters along the time axis to consider time-varying characteristics of speech;  FDY has been shown to improve sound event detection by dynamically adapting the filters along the frequency axis,  addressing the fact that the frequency dimension is not shift invariant. However, both TDY and FDY execute $K$ parallel convolutions before combining the results, which leads to considerable computational overhead during inference.

\textbf{Dynamic Activation Function:}
Less explored in comparison to CNN attention mechanisms and dynamic convolutions is the use of dynamic activation functions. Si et al.~\cite{si2018dynamic} dynamically adapt the threshold value of ReLU activations in an MLP network. Chen et al. extend this line of research by introducing dynamic ReLU (Dy-ReLU)~\cite{chen2020dynamic_relu}, which works as a dynamic and efficient version of Maxout~\cite{goodfellow2013maxout}. A hyperfunction, similar to Squeeze-and-Excitation~\cite{Hu18Squeeze}, is used to dynamically generate coefficients for $M$ linear mappings. After normalizing the coefficients (slopes and intercepts) to specific ranges, the element-wise maximum is applied across the $M$ different mappings. Most commonly, $M=2$ is used and linear mappings are spatially shared but differ across channels.  

\subsection{Pre-trained Audio Models}

Models in the audio domain are typically pre-trained in a supervised or self-supervised way on large-scale datasets, such as AudioSet~\cite{audioset2017Gemmeke}. AudioSet consists of around 2 million weakly labeled 10-second audio snippets downloaded from YouTube. The audio clips were manually annotated with 527 different event classes hierarchically sorted in an ontology.

PANNs~\cite{Kong20PANNs} introduced a series of AudioSet-pre-trained CNNs of varying complexities and architectures, which are widely used for downstream applications in audio-related domains such as sound event detection~\cite{Ronchini2022dcase_task4}, automated audio captioning~\cite{mei2022automated}, language-based audio retrieval~\cite{Xie2022dcase_task6b}, emotion recognition~\cite{luna2021multimodal}, or even optical fiber sensing~\cite{Tonami2023optical}. The prevalence of PANNs across many different downstream application areas underlines the community's interest in pre-trained audio models for end-to-end fine-tuning. The study presented in~\cite{Kong20PANNs} includes MobileNetV2~\cite{Sandler18MobileNetsV2} which shows a good performance-complexity trade-off but falls behind CNN14~\cite{Kong20PANNs} in terms of performance. Gong et al.~\cite{Gong21PSLA} improve over PANNs in terms of performance and complexity by using an EfficientNet-B2~\cite{Tan19EfficientNet} pre-trained on ImageNet~\cite{Deng09ImageNet}, balanced sampling, and label enhancement. ERANNs~\cite{Verbitskiy21ERANN} improve the performance on AudioSet further while controlling efficiency by temporal downsampling via strided convolutions. However, despite all these improvements, CNNs have been substantially outperformed by supervised~\cite{Gong21Ast, Koutini21Passt, Chen22HTS-AT, Gong22CMKD} and self-supervised~\cite{Huang22Masked, chen2022beats, Gong2022ssast} Transformer models.

Recently, it has been shown that the sharp increase in audio tagging performance on AudioSet achieved by Transformers can be exploited and transferred to efficient CNNs using Knowledge Distillation (KD)~\cite{Hinton2015distilling, Ba14KD}. In this context, Gong et al.~\cite{Gong22CMKD} found that Transformers and CNNs are good teachers for each other, improving the performance of both models, and Schmid et al.~\cite{Schmid22Efficient} use Transformer-to-CNN KD to match the performance of a PaSST~\cite{Koutini21Passt} Transformer model with a MobileNetV3 having only 6\% of the parameters and requiring 100 times fewer MACs. These efficient, pre-trained CNNs have been shown to extract high-quality audio representations~\cite{schmid2023low} and have a high potential to be fine-tuned for low-complexity on-device applications. In parallel to our work, which focuses on architectural improvements of efficient CNNs with dynamic components, Dinkel et al.~\cite{dinkel2023ced} recently improved the KD setup introduced in \cite{Schmid22Efficient} by using consistent ensemble distillation and an improved teacher model.

% \subsection{Efficient Pre-Trained Audio Models}

% \subsection{Efficient Audio Classification}

% Besides porting efficient CNNs from the vision to the audio domain~\cite{Kong20PANNs, Gong21PSLA, Schmid22Efficient, Gong22CMKD} and designing efficient architectures directly for audio specific tasks~\cite{kim2021broadcasted, Schmid2023cp-mobile, Lopez-Meyer21EfficientAudioEmbeddings, Verbitskiy21ERANN, huang2018aclnet}, techniques such are pruning~\cite{frankle2021pruingatinit, Liu2014brain}, quantization~\cite{hubara17quantized, benoit18qat} and knowledge distillation (KD)~\cite{Hinton2015distilling, Ba14KD} are commonly used. In particular, KD is a widespread technique and has contributed substantially to recent advancements in efficient audio classification~\cite{Gong22CMKD, schmid2023low, gao2022multi, choi2022temporal, Schmid22Efficient, Schmid2023cp-mobile, tripathi2023divide}. 

\section{Proposed dynamic model}
\label{sec:dynamic}

This section introduces the proposed dynamic model consisting of dynamic ReLU (Dy-ReLU)~\cite{chen2020dynamic_relu}, dynamic convolutions (Dy-Conv)~\cite{chen2020dynamic_conv}, and Coordinate Attention (CA)~\cite{hou2021coordinate}. This design decision is based on our belief that the three dynamic methods are complementary: Dy-Conv can extract noise-invariant features~\cite{zhang2020dynet}, CA detects important channels, time frames and frequency bins, and Dy-ReLU increases the model's expressiveness by applying a dynamic non-linear function. These dynamic components are integrated into efficient inverted residual blocks (IR blocks)~\cite{Sandler18MobileNetsV2}, as our focus is on creating efficient pre-trained audio models. In particular, we use the global network design of MobileNetV3-Large (MN)~\cite{Howard19MobileNetV3} and scale the models by network width using a width multiplier $\alpha$. MN is optimized towards latency and has shown to provide an excellent performance--complexity trade-off on AudioSet~\cite{Schmid22Efficient}.
% ~\cite{audioset2017Gemmeke}.

In the following, we describe our proposed dynamic IR block in a top-down manner. We start by reviewing the conventional IR block in Section~\ref{subsec:inverted_block}, and introduce the modifications that lead to the dynamic IR block in Section~\ref{subsec:dynamic_block}. We then zoom into the four central components of the dynamic IR block: the Context Generation Module (Section~\ref{subsec:cgm}), Dy-ReLU (Section ~\ref{subsec:dyrelu}), Dy-Conv (Section~\ref{subsec:dyconv}) and CA (Section~\ref{subsec:ca}). For all these additional components, we will identify the computationally most costly part in terms of multiply-accumulate operations (MACs) and compare it to the cost of convolutions in the conventional IR block.

\begin{figure}[t]
\centering
{\includegraphics[width=240pt]{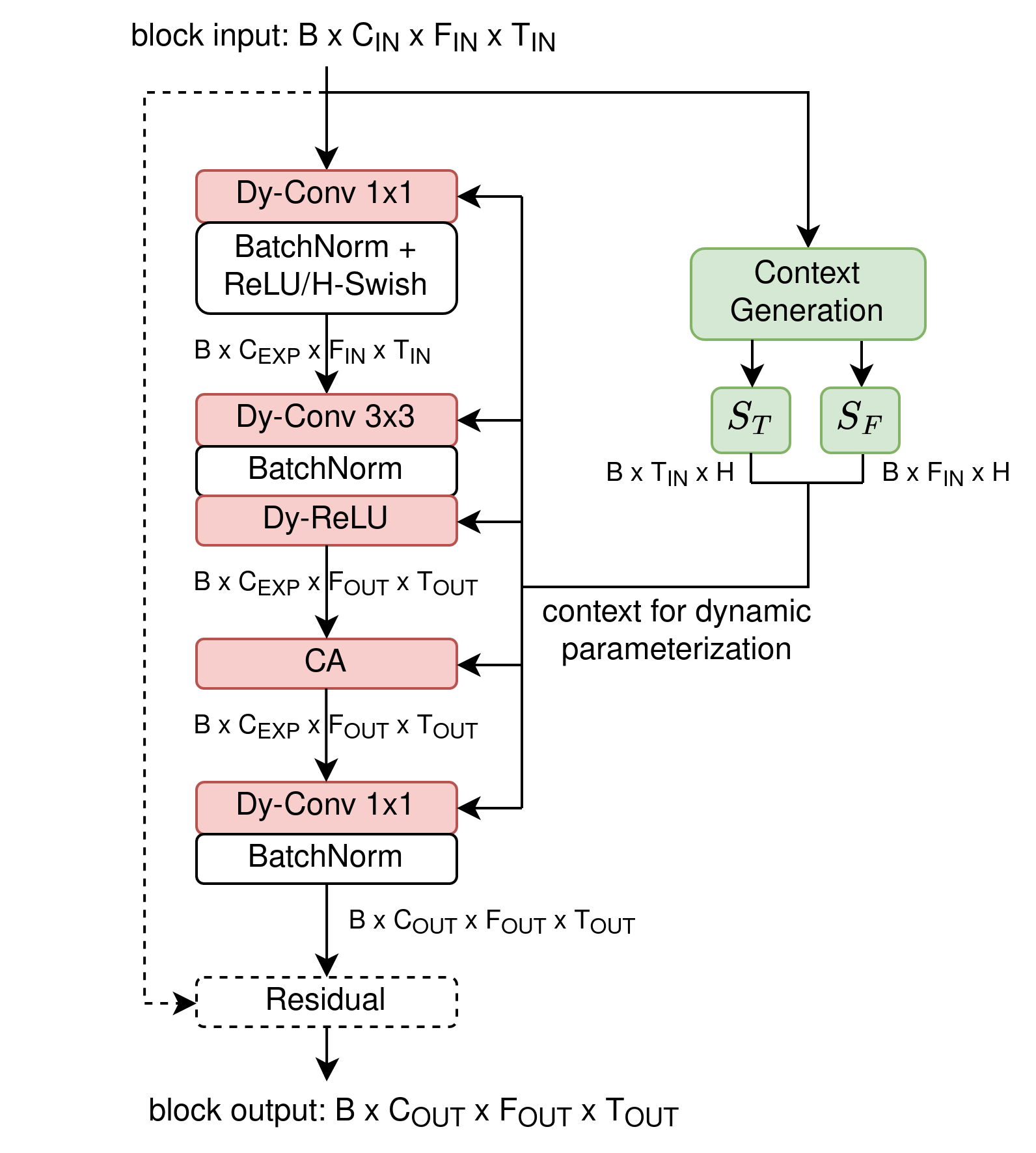}}
\caption{Dynamic inverted residual block: Starting from the conventional inverted residual (IR) block, all convolutions are replaced by dynamic convolutions (Dy-Conv)~\cite{chen2020dynamic_conv}; Coordinate Attention (CA)~\cite{hou2021coordinate} is used instead of Squeeze-and-Excitation~\cite{Hu18Squeeze}; and dynamic ReLU (Dy-ReLU)~\cite{chen2020dynamic_relu} replaces the non-linear activation function after the depthwise convolution. The Context Generation Module (CGM) operates on the block input and extracts embedded time and frequency sequences used to parameterize Dy-ReLU, Dy-Convs and CA. Dynamic components are depicted in red, and context generation is shown in green. The shape of the input feature map size is denoted in terms of \textit{batch size $\times$ channels $\times$ frequency bands $\times$ time frames}.}% caption command
\label{fig:dynamic_block}% label
\end{figure}

\subsection{Inverted Residual Block}
\label{subsec:inverted_block}

IR blocks~\cite{Sandler18MobileNetsV2} are constructed of (1) a pointwise channel expansion convolution projecting the number of channels from $C_{\mathit{IN}}$ to $C_{\mathit{EXP}}$ using 1x1 kernels, (2) a depthwise convolution operating on each of the $C_{\mathit{EXP}}$ channels independently, and (3) a pointwise projection convolution projecting the channels from $C_{\mathit{EXP}}$ to $C_{\mathit{OUT}}$ with 1x1 kernels. For most blocks, it holds that $C_{\mathit{IN}}=C_{\mathit{OUT}}$. However, transition blocks increase the number of channels, leading to $C_{\mathit{OUT}}>C_{\mathit{IN}}$. Each convolution is followed by batch normalization and ReLU activation, except for the linear bottleneck after the last convolution. The depthwise convolution can be strided to downsample the spatial dimensions. If the spatial dimensions and the number of channels match, a residual connection from block input to block output is used.

The pointwise convolutions are the computationally most expensive operations in an IR block. Specifically, given a block with $C_{\mathit{OUT}}$ output channels, $C_{\mathit{EXP}}$ channels in the expanded channel representation and spatial output dimensions of sizes $T_{\mathit{OUT}}$ and $F_{\mathit{OUT}}$, the final pointwise convolution performs $C_{\mathit{EXP}} * C_{\mathit{OUT}} * T_{\mathit{OUT}} * F_{\mathit{OUT}}$ MACs.

\begin{figure}[t]
\centering
{\includegraphics[width=160pt]{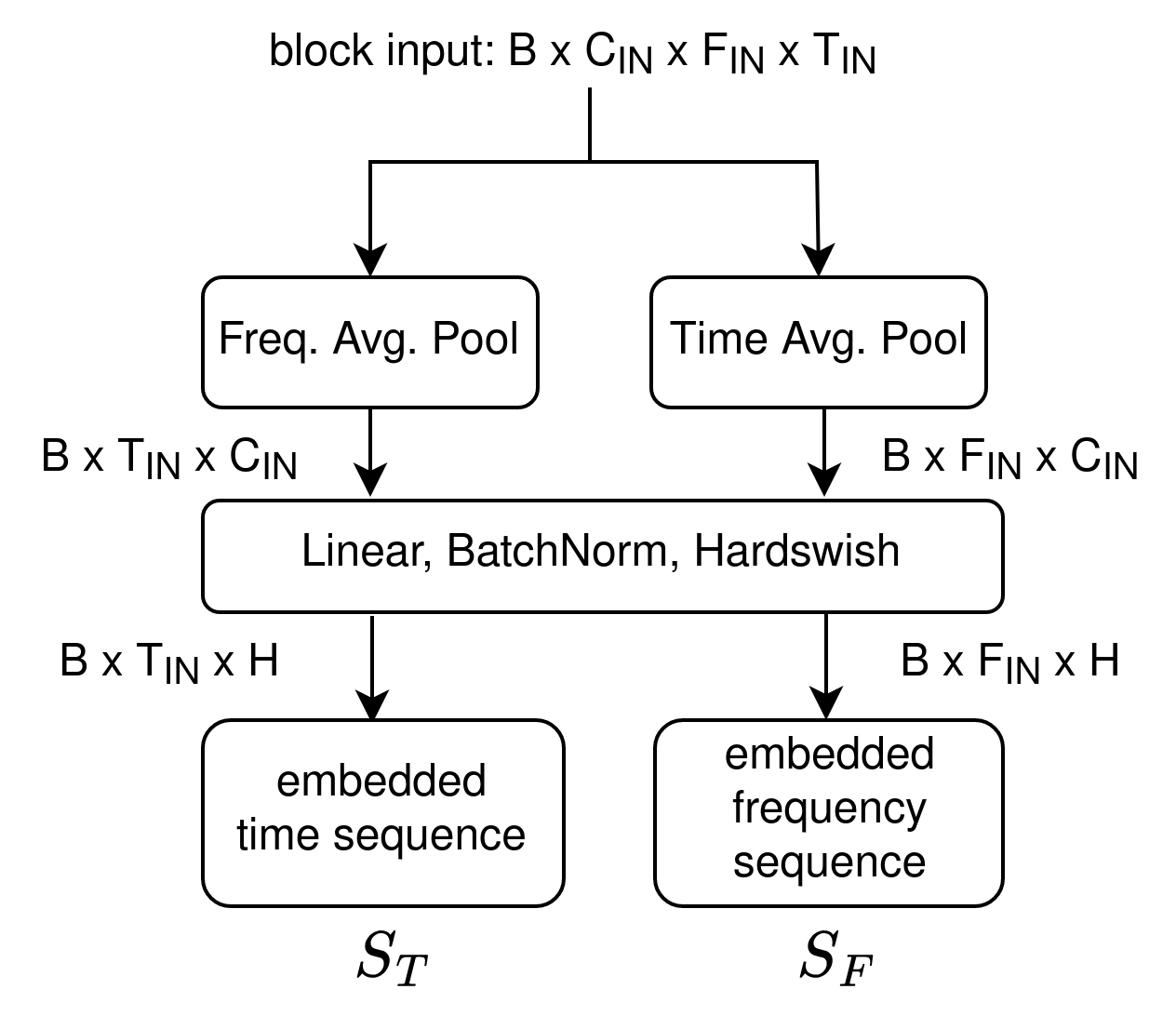}}
\caption{Context Generation Module (CGM): a zoom into the green parts of Fig.~\ref{fig:dynamic_block}. The CGM operates on the input of an IR block and outputs a time and frequency sequence embedded in a reduced channel dimension of size $H$. These sequences are used to parameterize Dy-ReLU~\cite{chen2020dynamic_relu} and Dy-Convs~\cite{chen2020dynamic_conv} and to compute the channel-time and channel-frequency recalibration weights for CA. The shape of the input feature map size is denoted in terms of \textit{batch size $\times$ channels $\times$ frequency bands $\times$ time frames}.}% caption command
\label{fig:context_generation}% label
\end{figure}

\subsection{Dynamic Inverted Residual Block}
\label{subsec:dynamic_block}

Starting from the conventional IR block, we apply the following modifications to integrate the dynamic components:
\begin{enumerate}
    \item \textbf{Conv $\xrightarrow{}$ Dy-Conv}: We replace all three  convolutions in the IR block with Dy-Conv~\cite{chen2020dynamic_conv}. 
    \item \textbf{ReLU $\xrightarrow[]{}$ Dy-ReLU}: The ReLU activation function after the depthwise convolution is replaced by Dy-ReLU~\cite{chen2020dynamic_relu}. As shown in Section~\ref{sec:ablation}, using additional Dy-ReLUs after the two pointwise convolutions does not yield further improvements.
    \item \textbf{Squeeze-and-Excitation $\xrightarrow[]{}$ CA}: Instead of SE~\cite{Hu18Squeeze}, CA~\cite{hou2021coordinate} is used as the attention mechanism. As will be shown in Section~\ref{sec:ablation}, this replacement yields substantial performance gains.
\end{enumerate}

Fig.~\ref{fig:dynamic_block} shows the overall structure of the dynamic IR block. All three types of dynamic components have in common that they require statistics extracted from the input sample for dynamic parameterization and reweighting. We share the computation of these statistics across all dynamic components in a common Context Generation Module (CGM), which will be explained in detail in Section~\ref{subsec:cgm}. The CGM operates on the input of the IR block and outputs an embedded time and frequency sequence, i.e., two separate lists of embeddings for time frames and frequency bins (depicted as green blocks in Fig.~\ref{fig:dynamic_block}). The following sections on Dy-ReLU (\ref{subsec:dyrelu}), Dy-Conv (\ref{subsec:dyconv}) and CA (\ref{subsec:ca}) will explain in detail how these sequences are processed for dynamic parameterization and reweighting.

\subsection{Context Generation Module}
\label{subsec:cgm}

The goal of the CGM is to collect informative statistics from an input sample that can be used to parameterize the dynamic components without creating substantial computational overhead. Fig.~\ref{fig:context_generation} depicts the details of this process. The CGM transforms the block input feature map into embedded time ($S_T$) and frequency ($S_F$) sequences. Inspired by the original CA module~\cite{hou2021coordinate}, introduced in the vision domain, the input feature map is pooled separately across the two spatial dimensions to retain positional information. The resulting time and frequency sequences are then processed by a shared transformation consisting of a linear layer, a batch-norm~\cite{ioffe2015batch} and hardswish activation~\cite{Howard19MobileNetV3}. The linear layer embeds the channel dimension of size $C_{\mathit{IN}}$ into a reduced space with $H$ dimensions. We set $H$ to a fraction of the expanded channel representation, such that $H=C_{\mathit{EXP}} / r$, where $r=4$ in our experiments.

The computationally most expensive operation in the CGM is the linear layer, which performs $C_{\mathit{IN}}*H*(T_{\mathit{IN}}+F_{\mathit{IN}})$ MACs. Compared to the first pointwise convolution in the IR block, which requires $C_{\mathit{IN}}*C_{\mathit{EXP}}*T_{\mathit{IN}}*F_{\mathit{IN}}$ MACs, the computational demand of the CGM is insignificant.

\subsection{Dy-ReLU}
\label{subsec:dyrelu}

Our specific implementation of dynamic ReLU is the spatially-shared and channel-wise Dy-ReLU-B introduced in~\cite{chen2020dynamic_relu}. Applying Dy-ReLU-B to a feature map requires predicting $M$ linear mappings for each channel. The output of Dy-ReLU-B is then computed by taking the elementwise maximum across the $M$ linear mappings. Specifically, given $x_c$, the input plane of the channel with index \textit{c}, and the predicted coefficients $\alpha_c^{m}$ (slope) and $\beta_c^{m}$ (intercept) for that channel, the output plane $y_c$ is calculated as follows:

\begin{equation}
\label{eq:dyrelu_max}
    \begin{split}
    y_c = \max_{1 \leq m \leq M}\{\alpha_c^{m} x_c + \beta_c^{m}\}
    \end{split}
\end{equation}

Dy-ReLU-B requires in total $2 * M * C$ dynamic coefficients, resulting from slope and intercept for $M$ linear mappings for each of the $C$ channels. 

We apply the transformation shown in Eq.~\ref{eq:dyrelu_coef} to predict the dynamic coefficients from the CGM output sequences $S_T$ and $S_F$. Firstly, the two sequences are concatenated and pooled over the sequence length dimension, resulting in a vector of size $H$. Secondly, a trainable linear transformation with parameters $W \in \mathbb{R}^{2*M*C_{\mathit{EXP}} \mathrm{x} H}$ and $b \in \mathbb{R}^{2*M*C_{\mathit{EXP}}}$ is used to predict the Dy-ReLU coefficients collected in the vector $C_{\mathrm{dyrelu}}$. Most commonly, \textit{M=2} linear mappings are used~\cite{chen2020dynamic_relu}, which is also the default value in our setup.

\begin{equation}
\label{eq:dyrelu_coef}
    \begin{split}
     C_{\mathrm{dyrelu}}= \mathrm{Pool}(\mathrm{Concat}[S_T, S_F]) W^T + b
    \end{split}
\end{equation}

The computational cost of Dy-ReLU is dominated by the calculation of elementwise linear mappings ($\alpha_c^m x_c + \beta_c^m$) as part of Eq.~\ref{eq:dyrelu_max}. The total cost of computing $M$ linear mappings is
$M*C_{\mathit{EXP}}*T_{\mathit{OUT}}*F_{\mathit{OUT}}$ MACs. Since $M=2$ is much smaller than the number of block output channels $C_{\mathit{OUT}}$, the cost of DyReLU is insignificant compared to the cost of the final pointwise convolution ($C_{\mathit{OUT}}*C_{\mathit{EXP}}*T_{\mathit{OUT}}*F_{\mathit{OUT}}$) as discussed in Section~\ref{subsec:inverted_block}. 

\begin{figure}[t]
\centering
{\includegraphics[width=220pt]{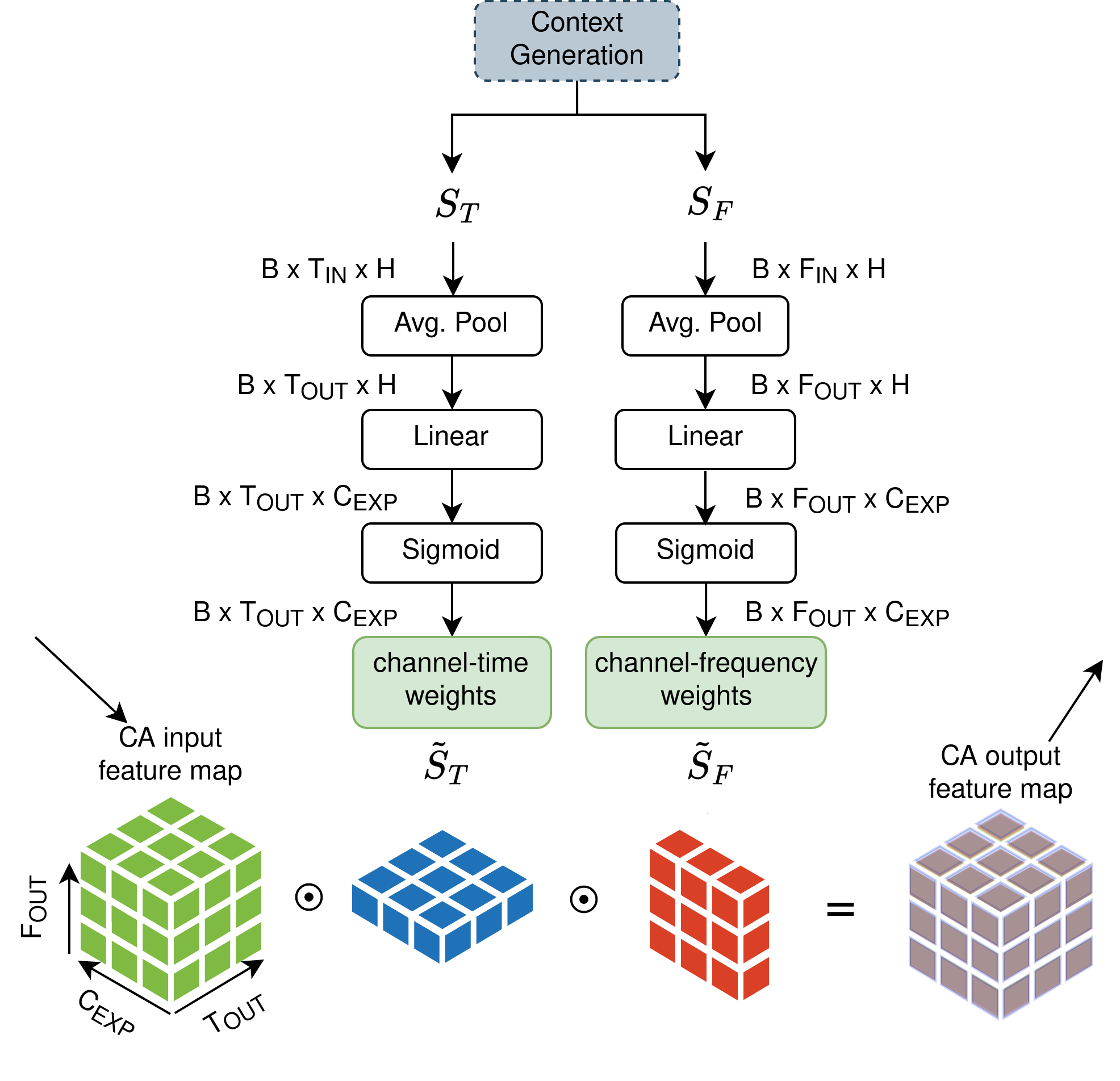}}
\caption{Coordinate Attention (CA): CA highlights important channels, frequency bins and time frames by recalibrating the feature map by element-wise multiplication with attention weights. It operates on the output sequences of the CGM and transforms them separately into the channel-time $\Tilde{S}_T$ and channel-frequency $\Tilde{S}_F$ attention weights. Average Pooling with a kernel size of 3 and a stride of 2 is used in case of a strided IR block. The linear transformation upsamples the number of channels from $H$ to $C_{\mathit{EXP}}$ to match the dimensionality of the feature map after the depthwise convolution.}% caption command
\label{fig:ca}% label
\end{figure}

\subsection{Dy-Conv}
\label{subsec:dyconv}

Our implementation of Dy-Conv is based on the dynamic convolution introduced in Chen et al.~\cite{chen2020dynamic_conv}. \textit{K} different kernels are kept in parallel and are aggregated based on predicted kernel attention weights $\Tilde{\alpha}_k$. The sum of the kernel attention weights is normalized to 1 using a softmax function with temperature scaling:

\begin{equation}
\label{eq:softmax_temp}
    \begin{split}
    \alpha_k = \frac{\mathrm{exp(}\Tilde{\alpha}_k/\tau\mathit{)}}{\sum_{k}^{K} \mathrm{exp(}\Tilde{\alpha}_k/\tau\mathit{)}}
    \end{split}
\end{equation} 

The aggregated dynamic kernel $W$ is then constructed as a weighted sum of the $K$ individual kernels: $W = \sum_{k}^{K} \alpha_k W_k$. Per default, we stick with the recommended settings in~\cite{chen2020dynamic_conv} and use $K=4$ and linearly anneal the temperature $\tau$ from 30 to 1 over the first epochs of training.

To obtain predictions from the CGM output sequences for the \textit{K} kernel attention weights, we apply the transformation shown in Eq.~\ref{eq:dyconv_coef}. The difference to the Dy-ReLU coefficient prediction is only in the shape of the trainable linear transformation and the resulting coefficients vector $C_{\mathrm{dyconv}}$. Specifically, we use $W \in \mathbb{R}^{K\mathrm{x} H}$ and $b \in \mathbb{R}^{K}$ resulting in a vector $C_{\mathrm{dyconv}}$ of size $K$.

\begin{equation}
\label{eq:dyconv_coef}
    \begin{split}
     C_{\mathrm{dyconv}}= \mathrm{Pool}(\mathrm{Concat}[S_T, S_F]) W^T + b
    \end{split}
\end{equation}

The computationally most expensive operation is the aggregation of the $K$ kernels requiring at most $K * C_{\mathit{EXP}} * C_{\mathit{OUT}}$ MACs for the final pointwise convolution. Since $K=4$ is much smaller than $T_{\mathit{OUT}} * F_{\mathit{OUT}}$, constructing the dynamic kernel is insignificant compared to the convolution itself.

\subsection{Coordinate Attention}
\label{subsec:ca}

The purpose of CA~\cite{hou2021coordinate} is to emphasize important spatial positions and channels by recalibrating a feature map with channel-time and channel-frequency weights. As depicted in Fig.~\ref{fig:ca}, CA takes as input the embedded time and frequency sequences as produced by the CGM, performs separate transformations per sequence, and outputs the respective attention weights. Eq.~\ref{eq:ca} shows the transformation from the embedded sequences $S_{\{T,F\}}$ to the respective attention weights $\Tilde{S}_{\{T,F\}}$. To match the dimensions of the feature map, an average pooling operation with kernel size 3 and a stride of 2 is applied in case of a strided depthwise convolution in the IR block. The result is processed by a trainable linear layer with parameters $W \in \mathbb{R}^{C_{\mathit{EXP}}\mathrm{x} H}$ and $b \in \mathbb{R}^{C_{\mathit{EXP}}}$ to match the channel dimension of the feature map. The final sigmoid function converts the resulting sequences into attention weights that are used to recalibrate the feature map via elementwise multiplications.

\begin{equation}
\label{eq:ca}
    \begin{split}
     \Tilde{S}_{\{T,F\}}= \mathrm{Sigmoid}(\mathrm{ Pool}(S_{\{T,F\}}) W^T + b)
    \end{split}
\end{equation}

The computationally most expensive operations in CA are the linear layers, which jointly perform $H * C_{\mathit{EXP}} * (T_{\mathit{OUT}} + F_{\mathit{OUT}})$ MACs. Compared to the final pointwise convolution in the IR block, which requires $C_{\mathit{OUT}} * C_{\mathit{EXP}} * T_{\mathit{OUT}} * F_{\mathit{OUT}}$ MACs, the computational demand of CA is negligible since $C_{\mathit{OUT}}$ $\approx$ H and the product of the spatial dimensions is typically much larger than their sum.

\section{Pre-Training on large-scale Audio Tagging}
\label{sec:pre-training}

In this section, we report the pre-training results of the introduced dynamic CNNs on the task of large-scale audio tagging on AudioSet~\cite{audioset2017Gemmeke}. That is, models need to assign one or multiple labels out of 527 classes to 10-second audio clips. Since AudioSet must be downloaded from YouTube, different proportions of the dataset can be successfully downloaded. In this regard, our setup is strictly comparable to the dataset used in \cite{Koutini21Passt} and \cite{Schmid22Efficient}. The proposed DyMN is scaled to three different complexities using width multipliers $\alpha \in \{0.4, 1, 2\}$. Adapting the width multiplier changes the number of channels in the IR blocks while keeping the total number of blocks constant. We denote the resulting models as DyMN small (DyMN-S), DyMN medium (DyMN-M) and DyMN large 
(DyMN-L). By default, we replace all 15 IR blocks with their dynamic counterparts; however, we will also discuss the effect of applying the dynamic IR blocks selectively in Section~\ref{sec:ablation}. The results will be compared in terms of parameter and computational efficiency to other models pre-trained on AudioSet. In particular, we are interested in a comparison to the non-dynamic counterpart of our proposed DyMNs, the efficient MNs used in \cite{Schmid22Efficient}.

\subsection{Pre-Training Setup}
\label{subsec:pre-training}

\subsubsection{Preprocessing and Augmentation} To pre-train our models on AudioSet, we match the preprocessing used in \cite{Koutini21Passt} and \cite{Schmid22Efficient}. We use mono audio with a sampling rate of 32 kHz and apply STFT with a window length of 25 ms and a hop size of 10 ms. Mel spectrograms are computed using a Mel filterbank with 128 frequency bins and the minimum and maximum frequency is randomly perturbed within a range of 10 Hz and 2 kHz, respectively. Mixup~\cite{Zhang18mixup} with a mixing coefficient of 0.3 is the only spectrogram-level data augmentation used since we are using offline KD as described in Section~\ref{subsec:offline_kd} and it has been shown that consistent KD is beneficial~\cite{Beyer21Consistent, Schmid22Efficient}. 

\subsubsection{Training} Models are trained for a total of 200 epochs and 100,000 samples are drawn at random without replacement from the full AudioSet in each epoch. We use a learning rate scheduler consisting of an exponential warmup phase until epoch 8, followed by a constant peak learning rate phase, 95 epochs of linear rampdown and 25 epochs fine-tuning with 1\% of the peak learning rate. The peak learning rates are set to \num{2e-3}, \num{1e-3} and \num{5e-4} for DyMN-S/M/L, respectively. We use the Adam optimizer~\cite{kingma2014adam} with a batch size of 120. We adopt the importance sampling strategy based on label frequency from~\cite{Koutini21Passt} to counter the long tail of infrequent classes. The results presented in this section are achieved by DyMNs pre-trained on ImageNet~\cite{Deng09ImageNet}, which has been shown to improve performance substantially~\cite{Gong21PSLA}. 

\subsubsection{Dynamic Component Settings} The temperature $\tau$ used in the context of Dy-Convs (see Eq.~\ref{eq:conv_mixture}) is linearly annealed from 30 to 1 in the first 30 epochs of training. The sequence embedding dimension $H$ as defined in the CGM (Fig.~\ref{fig:context_generation}) is set to $C_{\mathit{EXP}}/r$ with $r=4$. However, we additionally restrict its size between 32 and 128 to ensure that the sequences can capture enough information about the feature map in early layers  and do not get unnecessarily complex in the final layers. These bounds are scaled accordingly with the model's width multiplier $\alpha$.

\subsection{Offline Knowledge Distillation}
\label{subsec:offline_kd}

We copy the Transformer-to-CNN KD method introduced in \cite{Schmid22Efficient} to train our DyMNs. Specifically, the DyMNs act as student in the KD setup and optimize the loss given in Eq.~\ref{eq:ts_loss}. The loss is a weighted sum of label loss $L_\mathit{l}$ and distillation loss $L_\mathit{kd}$ traded off by a hyperparameter $\lambda$. $y$ denotes the AudioSet labels, $z_S$ and $z_T$ are the student and teacher logits, respectively, $\delta$ is the sigmoid activation function, and Binary-Cross-Entropy is applied for $L_\mathit{l}$ and $L_\mathit{kd}$. 

\begin{equation}
  \label{eq:ts_loss}
    Loss = \lambda L_\mathit{l}(\delta(z_S), y) + (1 - \lambda) L_\mathit{kd}(\delta(z_S), \delta(z_T))
\end{equation}

The teacher logits $z_T$ are constructed by ensembling the logits of 9 different Patchout FaSt Spectrogram Transformer (PaSST) models~\cite{Koutini21Passt} achieving a mean average precision (mAP) of 49.5 on the AudioSet evaluation set. Aligned with~\cite{Schmid22Efficient}, we pre-compute the ensemble logits for all recordings in the training set to speed up the training of the DyMN students and use $\lambda=0.1$ to emphasize the distillation loss. 

\subsection{Results on AudioSet}
\label{subsec:results_audioset}

\begin{figure}[t]
\centering
{\includegraphics[width=230pt]{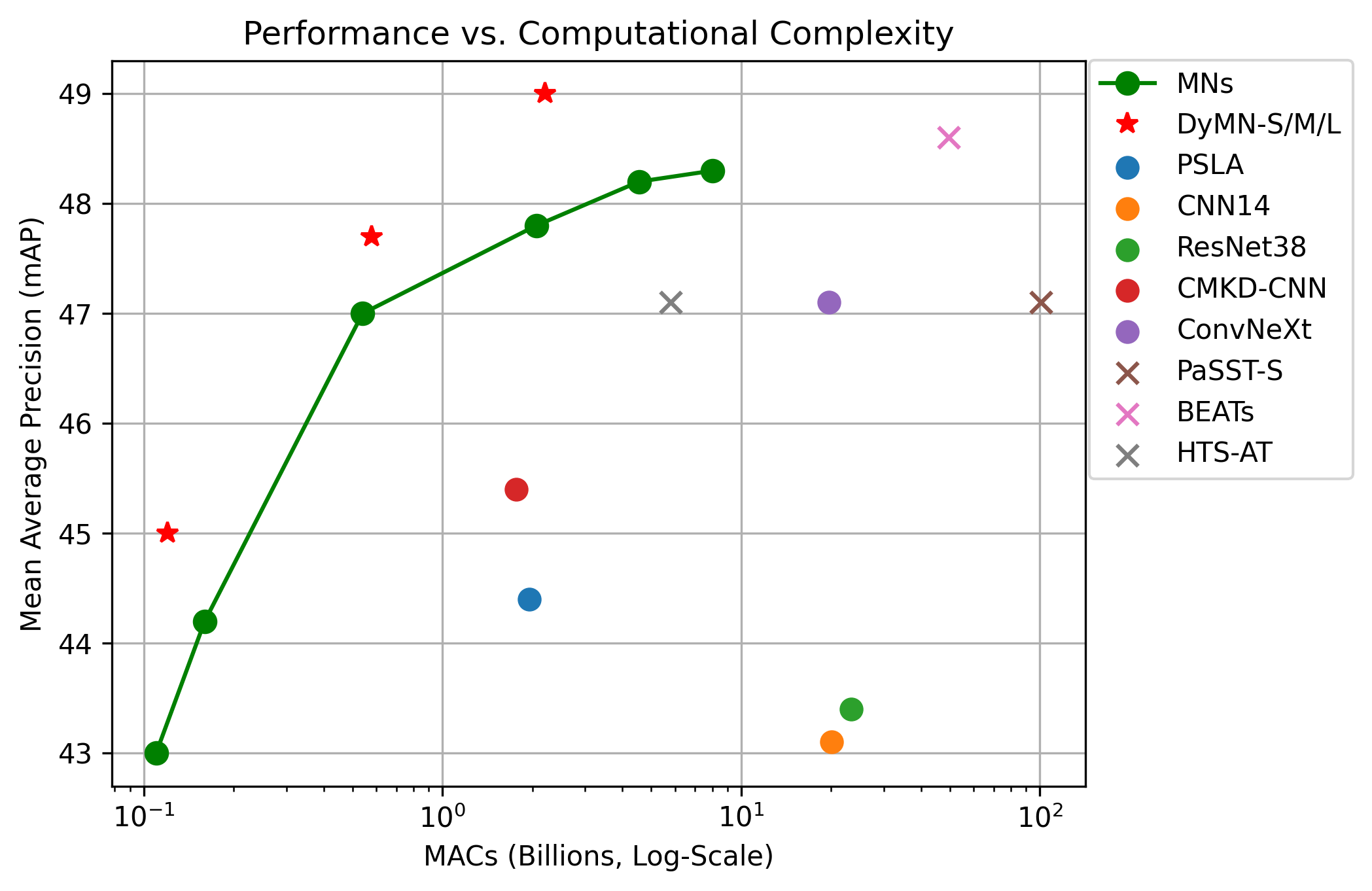}}
\caption{The plot compares the performance -- computational complexity trade-off across different single (i.e., non-ensemble) models on AudioSet. CNNs (MNs~\cite{Schmid22Efficient}, PSLA~\cite{Gong21PSLA}, CNN14~\cite{Kong20PANNs}, ResNet38~\cite{Kong20PANNs}, CMKD-CNN~\cite{Gong22CMKD}, ConvNeXt~\cite{pellegrini2023adapting}) are shown as circles, Transformer models (PaSST-S~\cite{Koutini21Passt}, BEATs~\cite{chen2022beats} and HTS-AT~\cite{Chen22HTS-AT}) are depicted as crosses, the width-scaled MobileNets introduced in~\cite{Schmid22Efficient} are connected by the green line (for better visibility), and our proposed DyMNs are denoted as red stars. The computational complexity is measured in terms of multiply-accumulate operations (MACs) plotted in log scale on the x-axis.}% caption command
\label{fig:computational_comp}% label
\end{figure}

In the following, we will compare the computational efficiency and the parameter efficiency across different models trained on AudioSet.
% ~\cite{audioset2017Gemmeke}.
In particular, the DyMNs are trained in the same setup as the MNs in \cite{Schmid22Efficient}, which permits a fair comparison in assessing the effect of the dynamic components. All results presented throughout this paper are averages of at least 3 independent runs and the last 10 epochs of training.

\begin{figure}[t]
\centering
{\includegraphics[width=230pt]{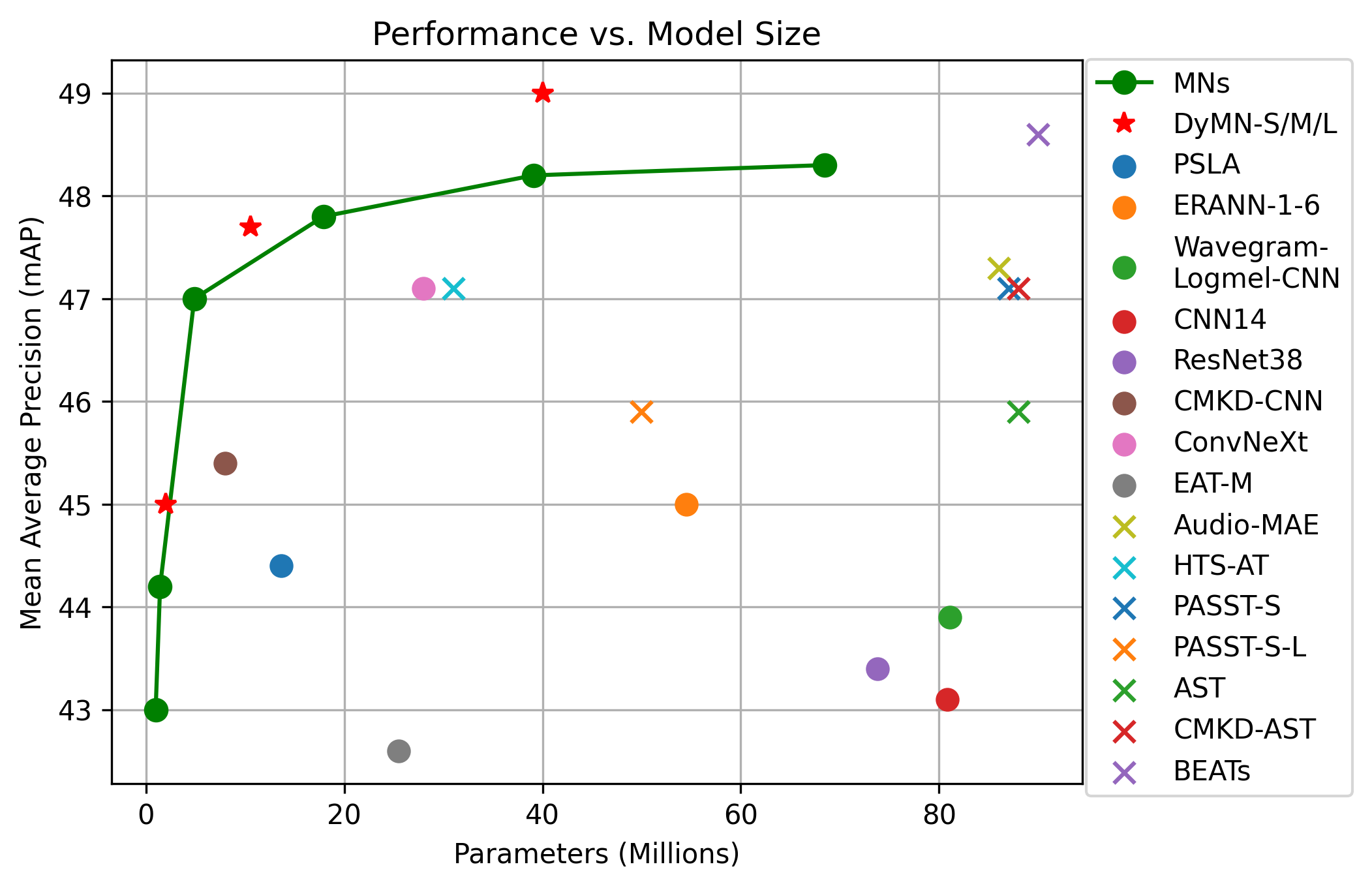}}
\caption{The plot compares the parameter-efficiency across multiple different single models on AudioSet\cite{audioset2017Gemmeke}. CNNs (MNs~\cite{Schmid22Efficient}, PSLA~\cite{Gong21PSLA}, ERANN~\cite{Verbitskiy21ERANN}, Wavegram-Logmel CNN~\cite{Kong20PANNs}, CNN14~\cite{Kong20PANNs}, ResNet38~\cite{Kong20PANNs}, CMKD-CNN~\cite{Gong22CMKD}, ConvNeXt~\cite{pellegrini2023adapting}, EAT-M~\cite{gazneli2022end}) are denoted as circles. Transformers (Audio-MAE~\cite{Huang22Masked}, HTS-AT~\cite{Chen22HTS-AT}, PaSST-S~\cite{Koutini21Passt}, PaSST-S-L~\cite{Koutini21Passt}, AST~\cite{Gong21Ast}, CMKD-AST~\cite{Gong22CMKD}, BEATs~\cite{chen2022beats}) are denoted as crosses. The green line connects the series of width-scaled MobileNets introduced in~\cite{Schmid22Efficient} and the red stars indicate our proposed DyMNs.}% caption command
\label{fig:parameter_comp}% label
\end{figure}

\subsubsection{Computational Complexity} 

The number of MACs is calculated on 10-second audio recordings using the model profiler contained in Microsoft's DeepSpeed framework~\cite{rasley2020deepspeed}. The results, plotting the performance of different models against the consumed MACs, are shown in Fig.~\ref{fig:computational_comp}. We compare our DyMNs (red stars) to other popular CNNs (circles) and Transformers (crosses) trained on AudioSet.
% ~\cite{audioset2017Gemmeke}.
The plot shows that DyMNs and MNs~\cite{Schmid22Efficient} (green line) trained in the Transformer-to-CNN KD setup~\cite{Schmid22Efficient} outperform other CNNs in terms of both prediction performance and computational efficiency. The diagram also confirms that the number of consumed MACs is indeed only marginally increased by the dynamic components  while they can boost the performance substantially. DyMN-S improves over its static counterpart MN with a matching width multiplier ($\alpha=0.4$) by almost 2 points in mAP. DyMN-M ($\alpha=1.0$) almost matches the performance of the MN with twice the width ($\alpha=2.0$), and DyMN-L ($\alpha=2.0$) outperforms even the largest MN ($\alpha=4.0$) while requiring approximately 4 times less MACs. The crosses depict the Transformer models PaSST-S~\cite{Koutini21Passt}, which is used as the teacher for MNs and DyMNs;\footnote{More precisely: we show here one single Transformer model from the teacher ensemble, in order to only have comparable single models in the plot. The 9-model PaSST ensemble would figure at $(6.4*10^{2},49.5)$ in Fig.~\ref{fig:computational_comp}; in Fig.~\ref{fig:parameter_comp}, it would completely distort the plot, with a parameter complexity of 775.3 million.} BEATs~\cite{chen2022beats}, which achieves the best performance on AudioSet; and HTS-AT~\cite{Chen22HTS-AT}, a particularly efficient implementation based on the Swin Transformer~\cite{liu2021swin}. While both DyMNs and MNs can outperform a single-teacher model, the PaSST ensemble teacher performance of 49.5 mAP is not reached. However, DyMN-L outperforms even the best-performing Transformer model BEATs~\cite{chen2022beats} while requiring less than 5\% of its MACs.

\subsubsection{Parameter Complexity} The parameter efficiency on AudioSet is compared across different CNNs (circles) and Transformers (crosses) in Fig.~\ref{fig:parameter_comp}. Both MNs and DyMNs outperform a variety of Transformers and CNNs in terms of the performance--complexity trade-off. BEATs~\cite{chen2022beats} is the only model that outperforms the best MNs, but DyMN-L achieves a higher mAP with less than half the number of parameters. The introduced DyMNs outperform the MNs in terms of parameter efficiency, although the dynamic convolutions require more than $K$ times as many parameters as conventional convolution layers and the fully-connected layers predicting the Dy-ReLU activation coefficients are nonnegligible in terms of parameters. However, scaling the width of a network by a factor $\alpha$ increases the number of parameters approximately by $\alpha^2$. This shows that introducing dynamic IR blocks to MNs instead of scaling by network width can be more efficient also in terms of parameters.

\section{Experiments on Downstream Tasks}
\label{sec:experiments}

The previous section has shown that the introduction of dynamic components to MNs increase the pre-training performance on large-scale AudioSet.
% ~\cite{audioset2017Gemmeke}.
However, the main question is if the performance gain during pre-training can be carried over to downstream tasks. We fine-tune AudioSet-pre-trained MNs and DyMNs on the tasks of polyphonic musical instrument recognition, environmental sound classification, sound event tagging, and acoustic scene classification, and compare their performance against each other, the pre-training teacher Transformer PaSST~\cite{Koutini21Passt} and the state of the art on the respective tasks. Furthermore, we share the same fine-tuning pipeline across all downstream tasks and only adapt the learning rate to show that no extensive hyperparameter tuning is required for high performance on downstream tasks.

\subsection{Tasks}

\subsubsection{Polyphonic Musical Instrument Recognition}

This task is to recognize all instruments present in an audio clip. It is based on the OpenMIC dataset~\cite{humphrey2018openmic} which consists of 20,000 10-second audio clips. Each clip is annotated by multiple tags out of 20 different classes. The performance metric is mean average precision (mAP). The state of the art for this task is the Transformer model PaSST~\cite{Koutini21Passt}, which surpassed receptive-field-regularized CNNs~\cite{Koutini21Receptive} as the previous state-of-the-art method.  

\subsubsection{Environmental Sound Classification}

This task is to classify 5-second audio recordings into one out of 50 different classes. It is based on the ESC50~\cite{piczak2015esc} dataset consisting of 2,000 environmental sound recordings. The performance metric for this task is accuracy and we report the results in terms of averages of the 5 official folds~\cite{piczak2015esc}. The state-of-the-art model on this dataset is the Transformer BEATs~\cite{chen2022beats}; in terms of CNNs the fine-tuned audio encoder of CLAP~\cite{elizalde2023clap} has the lead. 

\subsubsection{Sound Event Tagging}

The FSD50K dataset~\cite{fonseca2021fsd50k} consists of 51,197 recordings that are annotated with 200 event classes taken from the AudioSet~\cite{audioset2017Gemmeke} ontology. It is the second-largest publicly available general-purpose sound event tagging dataset after AudioSet, consisting of 100 hours of audio. FSD50K is separated into training, validation and evaluation splits. We use the validation split to set up our fine-tuning pipeline that is shared across all 
models and downstream tasks. The performance is reported in terms of mAP on the evaluation set. The multi-modal giant-size ONE-PEACE~\cite{wang2023one} Transformer with 4B parameters achieves state-of-the-art results on this dataset. On the CNN side, the fine-tuned  audio encoder of CLAP~\cite{elizalde2023clap} achieves the highest mAP.

\subsubsection{Acoustic Scene Classification}

This task is to classify 10-second audio recordings into one out of ten different acoustic scenes. The TAU Urban Acoustic Scenes 2020 Mobile dataset~\cite{heittola2020acoustic} has been used in the DCASE 2020 challenge \mbox{Task 1} and consists of 13,965 recordings in the train set and 2,979 in the test set. The performance is measured in terms of accuracy. It is particularly difficult to find models that generalize well on this dataset since it is recorded with a limited number of microphones, some of which are completely unseen during training and cause a distribution shift at test time. PaSST-S~\cite{Koutini21Passt} is the state-of-the-art method on this dataset, outperforming the top CNN~\cite{suh2020designing} from the DCASE 2020 challenge~\cite{heittola2020acoustic}.

\begin{table}[t]
\caption{
 Results on the downstream tasks based on OpenMic~\cite{humphrey2018openmic}, ESC50~\cite{piczak2015esc}, FSD50K~\cite{fonseca2021fsd50k} and DCASE20~\cite{heittola2020acoustic}. OpenMic and FSD50K use mean average precision, ESC50 and DCASE20 use accuracy as the metric.
}
\begin{tabular}{l|cccc}
   \textbf{Model}  &  \multicolumn{4}{c}{\textbf{Task}} \\ 
   
   \midrule
   
   & OpenMic & ESC50 & FSD50K & DCASE20 \\
     
      \midrule
% PaSST-B & 83.7 & 96.3 & 64.9 & 76.3   \\
Baseline & 79.5~\cite{humphrey2018openmic} & 76.9~\cite{piczak2015esc} & 43.4~\cite{fonseca2021fsd50k} & 54.1~\cite{heittola2020acoustic} \\
SOTA & 84.3~\cite{Koutini21Passt} & 98.1~\cite{chen2022beats} & 69.7~\cite{wang2023one} & 76.6~\cite{morocutti2023device} \\
SOTA CNN & 83.1~\cite{Koutini21Receptive} & 96.7~\cite{elizalde2023clap} & 58.6~\cite{elizalde2023clap} & 73.7~\cite{suh2020designing} \\
PaSST-S & 84.3 & 96.8 & 65.3 & 75.6 \\
\midrule
MN ($\alpha=0.4)$ & 82.0 & 93.2 & 59.7 & 69.5 \\
DyMN-S & 83.6 & 96.4 & 61.9 & 72.9 \\
%DyMN-S repl. se
% & 82.9 & 94.7 & 61.4 & 72.0  \\
\midrule
MN ($\alpha=1.0$) & 83.8 & 96.4 & 65.1 & 72.1 \\
DyMN-M & 84.4 & 96.4 & 64.2 & 73.6 \\
%DyMN-M repl. se
% & 84.1 & 96.2 & 64.2 & 73.1  \\
\midrule
MN ($\alpha=2.0$) & 84.7 & 96.9 & 65.4 & 73.2\\
DyMN-L & 85.5 & 97.4 & 65.5 & 75.7\\ \midrule
MN ($\alpha=3.0$) & 84.8 & 96.8 & 65.6 & 73.5 \\
\bottomrule
\end{tabular}
\label{tab:results_downstream}
\end{table}

\subsection{Fine-Tuning Setup}

For fine-tuning models on downstream tasks, the pre-processing of audio recordings applied in the pre-training stage, as discussed in Section~\ref{subsec:pre-training}, is matched. Except for the learning rate, we share the fine-tuning pipeline across all tasks and models. We use the Adam optimizer and train for 80 epochs. The learning rate schedule includes an exponential warmup phase for 10 epochs, followed by a linear rampdown for 65 epochs and 5 final epochs with 1\% of the peak learning rate. The temperature $\tau$ to compute the attention weights for Dy-Conv is fixed to 1. As for data augmentation, we randomly roll the waveform over time in a maximum range of $\pm$125ms. We use two-level mixup~\cite{Zhang18mixup}, both on the raw waveforms and on the spectrogram level, and the audio waveform is multiplied to change the gain by $\pm$7 dB. The min and max frequencies of the mel filterbank are randomly perturbed within ranges of 10 Hz and 2 kHz, respectively. Interestingly, a critical performance factor on the downstream tasks is the weight decay, which must be set to 0 to achieve high performance.

\subsection{Results}

The results on the four downstream tasks are given in Table~\ref{tab:results_downstream}. For each of the tasks, we specify the baseline performance (Baseline), global state of the art (SOTA), state of the art among CNNs (SOTA CNN), and the performance of the AudioSet teacher model PaSST-S~\cite{Koutini21Passt}. We also compare our proposed DyMNs to the MNs with matching width multiplier and add a MN with increased width of $\alpha=3.0$.  

\subsubsection{DyMNs vs. MNs} The DyMNs outperform the MNs with matching width across all tasks and model sizes, with the exception of $\alpha=1.0$ on FSD50K. DyMN-L even outperforms MN ($\alpha=3.0$) on OpenMic, ESC-50 and DCASE20 while being on par on FSD50K. These results underline that dynamic components can increase channel efficiency and generalization performance. 

\subsubsection{DyMNs vs. PaSST} DyMN-L outperforms the pre-training teacher model PaSST-S~\cite{Koutini21Passt} on all four downstream tasks while requiring less than half of its parameters and less than 3\% of its MACs for computing the predictions for a 10-second audio recording. DyMN-M achieves comparable performance on OpenMic and ESC-50, being 8 times smaller and requiring less than 1\% of the number of MACs compared to PaSST.

\subsubsection{DyMNs vs. SOTA CNN} DyMN-L beats the state-of-the-art CNN performance on all four downstream tasks. On FSD50K even the most lightweight models DyMN-S and MN ($\alpha=0.4$) outperform the top CNNs such as PSLA~\cite{Gong21PSLA} (56.7), CMKD-CNN~\cite{Gong22CMKD} (58.2) or CLAP~\cite{elizalde2023clap} (58.6). While CMKD-CNN is the smallest of the aforementioned CNNs with 8M parameters, MN ($\alpha=0.4$) and DyMN-S are below 1M and 2M parameters, respectively.

\subsubsection{DyMNs vs. SOTA}
On OpenMic, DyMN-M and DyMN-L outperform the state-of-the-art performance held by the Transformer model PaSST~\cite{Koutini21Passt}. On ESC-50, DyMN-L lags slightly behind the top method BEATs~\cite{chen2022beats} but outperforms other recent Transformers such as AST~\cite{Gong21Ast} (95.7), PaSST~\cite{Koutini21Passt} (96.8) or HTS-AT~\cite{Chen22HTS-AT} (97.0). However, DyMN-L is much more lightweight compared to BEATs, having less than half of its parameters and less than 5\% of its MACs. On FSD50K, besides the giant-size model ONE-PEACE~\cite{wang2023one} with 4B parameters, DyMN-L outperforms other recent Transformers such as PaSST~\cite{Koutini21Passt} (65.3) and CMKD-AST~\cite{Gong22CMKD} (61.7) and on DCASE20 DyMN-L lags only behind a specific version of PaSST (PaSST-B) that uses no patchout (76.6).

\begin{table}[t!]

\centering
\begin{minipage}[t]{\columnwidth}
\caption{Importance of Dy-ReLU, Dy-Conv and CA in the proposed dynamic IR block. \textbf{--} denotes that the respective dynamic components are removed from \textbf{Proposed DyMN}.}
\label{tab:abl_dyn_comp}
\centering 
\begin{tabular}{@{}lcc@{}}
\toprule
\textit{Dynamic Components} & \textbf{mAP} & \textbf{Diff.}  \\ \midrule
\textbf{Proposed DyMN} & $\textbf{47.5}$ & \textit{Ref. Val.} \\
MN Baseline~\cite{Schmid22Efficient} & 45.8  & -1.7 \\
MN Static (no SE) & 44.4  & -3.1 \\
\midrule
\textbf{--} CA & 46.9 & -0.6  \\ 
\textbf{--} Dy-ReLU & 46.9 & -0.6   \\ 
\textbf{--} Dy-Conv & 47.2 & -0.3  \\ \midrule
\textbf{--} CA, Dy-ReLU & 46.1 & -1.4   \\ 
\textbf{--} CA, Dy-Conv & 46.6 & -0.9  \\ 
\textbf{--} Dy-ReLU, Dy-Conv & 46.7 & -0.8 \\ 
 \bottomrule
\end{tabular}
\end{minipage}

\bigskip

\centering
\begin{minipage}[t]{\columnwidth}
\caption{Effect of selective application of dynamic IR blocks.}
\label{tab:abl_selective_blocks}

\centering
\begin{tabular}{@{}lcc@{}}
\toprule
\textit{Dynamic Blocks} & \textbf{mAP} & \textbf{Diff.}  \\ \midrule
\textbf{Proposed DyMN} & $\textbf{47.5}$ & \textit{Ref. Val.} \\
MN Baseline~\cite{Schmid22Efficient} & 45.8  & -1.7 \\
\midrule
First 5 blocks dynamic & 46.3 &  -1.2 \\ 
Mid 5 blocks dynamic & 46.2 &  -1.3  \\ 
Last 5 blocks dynamic & 46.9 &  -0.6 \\ 
Replace SE & 47.1 & -0.4   \\ 
 \bottomrule
\end{tabular}
\end{minipage}

\end{table}

\section{Systematic Configuration Study}
\label{sec:ablation}

The purpose of this section is to justify the design decisions that led to the final dynamic IR block presented in Section~\ref{sec:dynamic}, and to show that the proposed variant has turned out to be beneficial across a variety of other configurations.

In the following, we present configuration studies for the proposed dynamic IR block in Section~\ref{subsec:abl_block}. We then delve into the details of the individual dynamic components in Sections~\ref{subsec:abl_att}, \ref{subsec:abl_dyconv} and \ref{subsec:abl_dyrelu}, followed by a study on different context generation variants in Section~\ref{subsec:abl_cgm}. All experiments are conducted on AudioSet~\cite{audioset2017Gemmeke} using DyMN-M and MN ($\alpha=1.0$) without ImageNet~\cite{Deng09ImageNet} pre-training. In the following tables, the default values in our setup are indicated in bold.

\subsection{Dynamic IR block}
\label{subsec:abl_block}

In this section, we perform a configuration study based on the proposed dynamic IR block. We investigate the effect of the individual dynamic components, the impact of applying the dynamic IR blocks selectively, and the effect of applying Dy-Conv and Dy-ReLU at different positions in the block.

\subsubsection{Importance of Dynamic Components}

Table~\ref{tab:abl_dyn_comp} presents the results for the proposed DyMN, the MobileNetV3~\cite{Howard19MobileNetV3} Baseline (MN Baseline) from~\cite{Schmid22Efficient}, a fully static MN with no Squeeze-and-Excitation~\cite{Hu18Squeeze} (MN Static) and all other combinations of the three dynamic components in DyMNs.

The results show that dynamic input-dependent processing is important. The proposed DyMN improves the performance by 7\% over the static MN. While Dy-ReLU and CA are of equal importance, Dy-Conv leads to the smallest improvements. However, all three dynamic components are beneficial for the overall performance and improve over MN Baseline. 

\begin{table}[t!]

\centering
\begin{minipage}[t]{\columnwidth}
\caption{Effect of varying the position of Dy-ReLU and Dy-Convs in the dynamic IR block.}
\label{tab:abl_dyrelu_dyconv_pos}

\centering
\begin{tabular}{@{}lcc@{}}
\toprule
\textit{Dy-Conv/Dy-ReLU Pos.} & \textbf{mAP} & \textbf{Diff.}  \\ \midrule
\textbf{Proposed DyMN} & $\textbf{47.5}$ & \textit{Ref. Val.} \\
\midrule
Dy-Conv Pos. 2 & 47.0 & -0.5  \\ \midrule
Dy-ReLU Pos. 1 & 47.2 & -0.3 \\ 
Dy-ReLU Pos. 2 + 1 & 47.4 & -0.1  \\ 
Dy-ReLU Pos. 2 + 3 & 47.2 &  -0.3 \\ 
 \bottomrule
\end{tabular}
\end{minipage}

\bigskip

\centering
\begin{minipage}[t]{\columnwidth}
\caption{MobileNetV3 with different attention mechanisms integrated in all IR blocks before the final pointwise convolution.} 
\label{tab:abl_cnn_att}

\centering
\begin{tabular}{lcc}
\toprule
\textit{Attention Method}
 & \textbf{mAP}
 & \textbf{Diff.} \\ 
\midrule
MN Static & 44.4  & \textit{Ref. Val.} \\
\midrule
\textbf{MN w. CA} & \textbf{46.7} & +2.3 \\
MN w. TA & 46.1 & +1.7 \\
MN w. SRM & 45.9 &  +1.5 \\
MN w. GRN & 45.3 & +0.9 \\
MN w. SE & 45.1  & +0.7 \\
MN w. CBAM & 44.6 & +0.2 \\
MN w. GC & 44.2 & -0.2 \\
\bottomrule
\end{tabular}
\end{minipage}

\end{table} 

\subsubsection{Selectively applying the dynamic IR block}

The purpose of this experiment, with results summarized in Table~\ref{tab:abl_selective_blocks}, is to determine at which positions in the model the dynamic blocks have the highest impact. MN has in total 15 IR blocks, all of which are replaced by dynamic IR blocks in the proposed DyMN. In this study, we replace only the first, middle and last 5 blocks in the MN with dynamic IR blocks and keep conventional IR blocks at the remaining positions. Additionally, the setting \textit{Replace SE} uses the dynamic IR block only at the positions at which the original MN uses SE, resulting in 8 out of the 15 IR blocks being dynamic.

Table~\ref{tab:abl_selective_blocks} shows that dynamic blocks are beneficial at different positions in the model. Each selective variant outperforms the MN Baseline from~\cite{Schmid22Efficient}. The best choice among the versions applying the 5 dynamic blocks is to make the last 5 blocks dynamic. Replacing only SE blocks with dynamic blocks comes closest to the fully dynamic model in terms of performance and can be seen as a lightweight alternative.

\subsubsection{Effects of Dy-Conv and Dy-ReLU positions}

The proposed dynamic IR block replaces all convolution layers with Dy-Conv and uses Dy-ReLU only after the depthwise convolution.  Table~\ref{tab:abl_dyrelu_dyconv_pos} shows the results for applying Dy-Conv and Dy-ReLU at alternative positions. Pos. 1, 2 and 3 describe the first, second and third convolution in the dynamic IR block (shown in Fig.~\ref{fig:dynamic_block}) and the activation functions that follow them. In case of Dy-ReLU at Pos. 3, we add an additional Dy-ReLU after the final pointwise convolution. 

The results show that replacing all convolution layers with Dy-Conv is beneficial and the proposed Dy-ReLU variant, where we have a single Dy-ReLU at Pos. 2, achieves the best performance. In particular, adding additional Dy-ReLUs does not improve results further.

\begin{table}[t!]

\centering
\begin{minipage}[t]{\columnwidth}
\caption{Effect of applying CA selectively for channel-time and channel-frequency recalibration.}
\label{tab:abl_ca}

\centering
\begin{tabular}{@{}lcc@{}}
\toprule
\textit{CA Recalibration} & \textbf{mAP} & \textbf{Diff.}  \\ \midrule
\textbf{Proposed DyMN} & $\textbf{47.5}$ & \textit{Ref. Val.} \\
\midrule
Only channel-frequency & 47.2  & -0.3  \\ 
Only channel-time & 47.2 &  -0.3  \\ 
 \bottomrule
\end{tabular}
\end{minipage}

\bigskip

\centering
\begin{minipage}[t]{\columnwidth}
\caption{Number of dynamic kernels $K$ in Dy-Conv}
\label{tab:abl_dyn_kernels_k}

\centering
\begin{tabular}{@{}lcc@{}}
\toprule
$K$ & \textbf{mAP} & \textbf{Diff.} \\ \midrule
2 & 47.4 & -0.1\\
\textbf{4} & \textbf{47.5} & \textit{Ref. Val.} \\
6 & 47.5 & 0.0\\ 
 \bottomrule
\end{tabular}
\end{minipage}

\end{table}

\subsection{Attention Mechanism}
\label{subsec:abl_att}

This section presents a study on the choice of attention mechanism and the impact of channel-frequency and channel-time recalibration in CA. 

\subsubsection{Choice of Attention Mechanism} 

Table~\ref{tab:abl_cnn_att} shows the results for integrating different popular attention mechanisms (CA~\cite{hou2021coordinate}, TA~\cite{misra2021triplet}, SRM~\cite{lee2019srm}, GRN~\cite{woo2023convnextv2}, SE~\cite{Hu18Squeeze}, CBAM~\cite{woo2018cbam}, and GC~\cite{cao2019gcnet}) into MN. All attention mechanisms are integrated before the final pointwise convolution into all 15 IR blocks.

While a number of different attention methods are capable of achieving substantial improvements over the static MN with no attention mechanism, CA leads to the largest improvement and is therefore the attention mechanism of choice for our proposed dynamic IR block.

\subsubsection{Channel-Time and Channel-Frequency Recalibration in CA} 

CA performs recalibration of the feature map with channel-time and channel-frequency attention weights. The results given in Table~\ref{tab:abl_ca} assess the importance of these two recalibration steps. While the channel-frequency and channel-time weights are equally important, using both of them leads to the best results.

\subsection{Dy-Conv}
\label{subsec:abl_dyconv}

In this section, the impact of the two hyperparameters of Dy-Conv, the number of kernels $K$ and the temperature $\tau$ is assessed.

\subsubsection{Number of dynamic kernels $K$}

The number of kernels $K$ specifies how many different kernels are aggregated in each Dy-Conv layer. Table~\ref{tab:abl_dyn_kernels_k} shows the results for $K \in \{2,4,6\}$. The performance improves only marginally from $K=2$ to $K=4$ kernels and plateaus for larger values of $K$.

\subsubsection{Temperate $\tau$}

The temperature $\tau$ affects the computation of kernel attention weights as shown in Eq.~\ref{eq:softmax_temp}. 
Aligned with~\cite{chen2020dynamic_conv}, by default we use a temperature schedule for $\tau$ and anneal it from 30 to 1 over the first 30 epochs of training. This ensures near-uniform attention weights in the first epochs to properly update all kernels. Table~\ref{tab:abl_dyconv_tau} compares the temperature schedule to the results of using a constant temperature ($\tau \in \{1, 10, 30\}$). The results show that the performance is stable across different constant temperature values in our setup. $\tau=10$ achieves the same performance as the temperature schedule. However, keeping the temperature constant and setting it to non-optimal values ($\tau=1$ or $\tau=30$) leads to a slight performance decrease, underlining the advantage of using a temperature schedule.

\begin{table}[t!]

\centering
\begin{minipage}[t]{\columnwidth}
\caption{Temperature $\tau$ for computing Dy-Conv attention weights}
\label{tab:abl_dyconv_tau}

\centering
\begin{tabular}{@{}lcc@{}}
\toprule
\textit{Temperature} & \textbf{mAP} & \textbf{Diff.} \\ \midrule
\textbf{$\tau$ annealing} & 47.5 & \textit{Ref. Val.}\\ \midrule
$\tau=1$ & 47.3 & -0.2 \\
$\tau=10$ & 47.5 & 0.0 \\
$\tau=30$ & 47.4 & -0.1 \\
 \bottomrule
\end{tabular}
\end{minipage}

\bigskip

\centering
\begin{minipage}[t]{\columnwidth}
\caption{Number of linear mappings $M$ in Dy-ReLU.}
\label{tab:abl_dyrelu_m}

\centering
\begin{tabular}{@{}lcc@{}}
\toprule
\textit{M} & \textbf{mAP} & \textbf{Diff.} \\ \midrule
1 & 47.0 & -0.5 \\
\textbf{2} & 47.5 & \textit{Ref. Val.}\\
3 & 47.5 & 0.0\\
 \bottomrule
\end{tabular}
\end{minipage}

\end{table}

\subsection{Dy-ReLU}
\label{subsec:abl_dyrelu}

An important hyperparameter of Dy-ReLU is the number of linear mappings $M$ that the max operation, shown in Eq.~\ref{eq:dyrelu_max}, acts on. The results for $M \in \{1,2,3\}$ are shown in Table~\ref{tab:abl_dyrelu_m}. $M=1$ results in a dynamic linear function while $M=2$ and $M=3$ are dynamic non-linear functions. The non-linear functions outperform the linear function and aligned with the findings in~\cite{chen2020dynamic_relu}, $M=2$ and $M=3$ achieve a similar performance.

\begin{table}[t!]

\centering
\begin{minipage}[t]{\columnwidth}
\caption{Different settings for context generation.}
\label{tab:abl_cgm}

\centering
\begin{tabular}{@{}lcc@{}}
\toprule
\textit{Context Method} & \textbf{mAP} & \textbf{Diff.}  \\ \midrule
\textbf{Proposed DyMN} & $\textbf{47.5}$ & \textit{Ref. Val.} \\
\midrule
no shared context & 47.2 & -0.3 \\ 
no shared seq. parameters & 47.4 & -0.1 \\ 
concat pooled seq. & 47.4 & -0.1 \\ 
 \bottomrule
\end{tabular}
\end{minipage}

\end{table}

\begin{table}[t!]

\centering
\begin{minipage}[t]{\columnwidth}
\caption{Varying the sequence embedding size $H$.}
\label{tab:abl_seq_size}

\centering
\begin{tabular}{@{}lcc@{}}
\toprule
\textit{Context Size} & \textbf{mAP} & \textbf{Diff.}  \\ \midrule
$r=4$ & \textbf{47.5} & \textit{Ref. Val.}  \\ 
$r=8$ & 47.3 & -0.2  \\
$r=16$ & 47.1 & -0.4  \\ \midrule
$H_{\mathit{MIN}}=16$ & 47.3 & -0.2 \\ 
$H_{\mathit{MIN}}=32$ & \textbf{47.5} & \textit{Ref. Val.} \\ 
$H_{\mathit{MIN}}=64$ & 47.4 & -0.1 \\ \midrule
$H_{\mathit{MAX}}=64$ & 47.2 & -0.3 \\ 
$H_{\mathit{MAX}}=128$ & \textbf{47.5} & \textit{Ref. Val.} \\ 
$H_{\mathit{MAX}}=256$ & 47.5 &  0.0 \\
 \bottomrule
\end{tabular}
\end{minipage}

\end{table}

\subsection{Context Generation}
\label{subsec:abl_cgm}

In this section, different variants of context generation are studied. In particular, architectural variants are discussed in Section~\ref{subsubsec:cgm_variants} and modifications of the context size $H$ are investigated in Section~\ref{subsubsec:cgm_H}.

\subsubsection{Different architectural variants for context generation}
\label{subsubsec:cgm_variants}

Table~\ref{tab:abl_cgm} contains results for the following modifications of the context generation process:
\begin{itemize}
    \item \textit{no shared context}: Refers to a setting in which Dy-Conv and Dy-ReLU extract their own context by GAP and a learnable non-linear transformation, as originally proposed in \cite{chen2020dynamic_conv} and \cite{chen2020dynamic_relu}, respectively. In this case, Dy-Conv and Dy-ReLU do not make use of the CGM output sequences. This experiment tests whether the shared CGM is capable of extracting a sufficiently rich global context that can be used to parameterize all dynamic components in a block.
    \item \textit{no shared seq. parameters}: Indicates that, in contrast to the CGM setup in Fig.~\ref{fig:context_generation}, the linear layer and batchnorm parameters are not shared across the sequences. Instead, two sets of parameters are learned to transform the time and frequency sequences. The motivation for this experiment is to decouple transformations involving time and frequency information, which, in contrast to the height and width of an image, encode different physical properties.
    \item \textit{concat pooled seq.}: Describes a setting for which Equations \ref{eq:dyrelu_coef} and \ref{eq:dyconv_coef} are modified. Specifically, the sequences $S_T$ and $S_F$ are pooled separately and the vectors of size $H$ are concatenated, resulting in a context vector of size $2*H$. Aligned with the motivation of the last experiment, with this experiment, we try to avoid mixing time and frequency information.
\end{itemize}

The results presented in Table~\ref{tab:abl_cgm} show that all of these modifications lead to a slight decrease in performance, despite all of them increasing the number of parameters. The proposed design of the context generation is based on the findings of these experiments; the CGM follows the feature encoding process used in CA~\cite{hou2021coordinate}, and the dynamic coefficients of Dy-ReLU and Dy-Conv are derived as defined in Eqs.~\ref{eq:dyrelu_coef} and \ref{eq:dyconv_coef}, respectively.

\subsubsection{Varying the sequence embedding size $H$}
\label{subsubsec:cgm_H}

As stated in Section~\ref{subsec:abl_cgm}, the embedding dimension for the time and frequency sequences, computed by the CGM, is defined as $H=C_{\mathit{EXP}}/r$. Additionally $H$ is clipped between a lower bound ($H_{\mathit{MIN}}=32$)  and an upper bound ($H_{\mathit{MAX}}=128$) that are scaled accordingly with the model's width $\alpha$. Table~\ref{tab:abl_seq_size} shows the results for different values of $r$, $H_{\mathit{MIN}}$ and $H_{\mathit{MAX}}$. The results indicate that reducing the sequence embedding dimension $H$ by either increasing $r$ or decreasing $H_{\mathit{MIN}}$ and $H_{\mathit{MAX}}$ leads to a decrease in mAP. This shows that a assigning a certain capacity to the global context is important for the dynamic components to exploit their full potential. However, increasing $H$ by increasing the lower and upper bounds ($H_{\mathit{MIN}}=64$ and $H_{\mathit{MAX}}=256$) does not yield further performance improvements and shows that the performance saturates after a certain context size is reached.

%\section{Limitations and Future Work}
%\label{sec:limitations}

%%% second tables column

\begin{table}[t!]
\centering
\caption{Inspecting Dynamic Components}
\label{tab:inspecting}
\begin{tabular}{@{}llcc@{}}
\toprule
\textit{Dynamic Component}&\textit{Inspection Method} & \textbf{mAP} & \textbf{Diff.}  \\ \midrule
&\textbf{Proposed DyMN} &
\textbf{47.8} & \textit{Ref. Val.} \\
&Random Predictions &
$0.005$ & -47.795 \\
\midrule
\parbox[t]{10mm}{\multirow{5}{*}{\rotatebox[origin=c]{0}{CA}}}
& context shuffle & 19.6 &  -28.2  \\
& channel shuffle & 2.5 &   -45.3 \\ 
& spatial shuffle & 12.5  &  -35.3  \\
& time shuffle & 35.7 &  -12.1  \\ 
& frequency shuffle & 22.0 &  -25.8  \\ 
\midrule
\parbox[t]{10mm}{\multirow{4}{*}{\rotatebox[origin=c]{0}{Dy-Conv}}} 
& context shuffle & 43.7 &  -4.1  \\
& attention shuffle & 43.4 &  -4.4 \\ 
& uniform attention & 45.3 &  -2.5 \\ 
& max attention & 39.5  &  -8.3 \\ 
\midrule
\parbox[t]{10mm}{\multirow{2}{*}{\rotatebox[origin=c]{0}{Dy-ReLU}}}
& context shuffle & 32.5 &  -15.3 \\ 
& channel shuffle & 3.5 &   -44.3 \\ 
 \bottomrule
\end{tabular}
\end{table}

\section{Inspecting the Dynamic Components}
\label{sec:inspecting}

In this section, we analyze the dynamic components and investigate whether the CA attention weights and predicted coefficients for Dy-ReLU and Dy-Conv are indeed input-dependent. In contrast to Section~\ref{sec:ablation}, the study conducted in this section is performed using a single well-trained model on AudioSet
% ~\cite{audioset2017Gemmeke} \gw{no need to cite these datasets again and again ...}
with ImageNet
% ~\cite{Deng09ImageNet} 
pre-training. In the following, we will discuss the dynamic behaviour of CA, Dy-Conv and Dy-ReLU in Sections~\ref{subsec:inspecting_ca}, \ref{subsec:inspecting_dy_conv} and \ref{subsec:inspecting_dy_relu}. All results are presented in Table~\ref{tab:inspecting}. The inspection method \textit{context shuffle} is shared across the three dynamic components and describes the following setting: The attention weights for CA and the coefficients for Dy-ReLU and Dy-Conv are applied to an input recording $\mathbf{x}$ but computed based on a different recording $\mathbf{\Tilde{x}}$. In particular, \textit{context shuffle} serves as the main analysis to determine whether the dynamic components perform input-dependent transformations.

\subsection{Coordinate Attention}
\label{subsec:inspecting_ca}

For \textit{context shuffle}, the performance decreases by more than 50\%, which clearly shows that the dynamic recalibration weights of CA are input-dependent. To assess the importance of channel, time and frequency attention, we randomly shuffle the recalibration weights along (1) the channel dimension (\textit{channel shuffle}), both spatial dimensions (\textit{spatial shuffle}), the time dimension (\textit{time shuffle}) and the frequency dimension (\textit{frequency shuffle}). The results in Table~\ref{tab:inspecting} show that the channel attention mechanism is the most important and attention over frequency bins is more important than attention over time frames. Surprisingly, \textit{channel shuffle} and \textit{spatial shuffle} lead to a more severe performance drop than \textit{context shuffle}, which indicates that, besides dynamic input-dependent processing, a shared prior across the samples is learned in CA.

\begin{figure*}[t]
\centering
{\includegraphics[width=400pt]{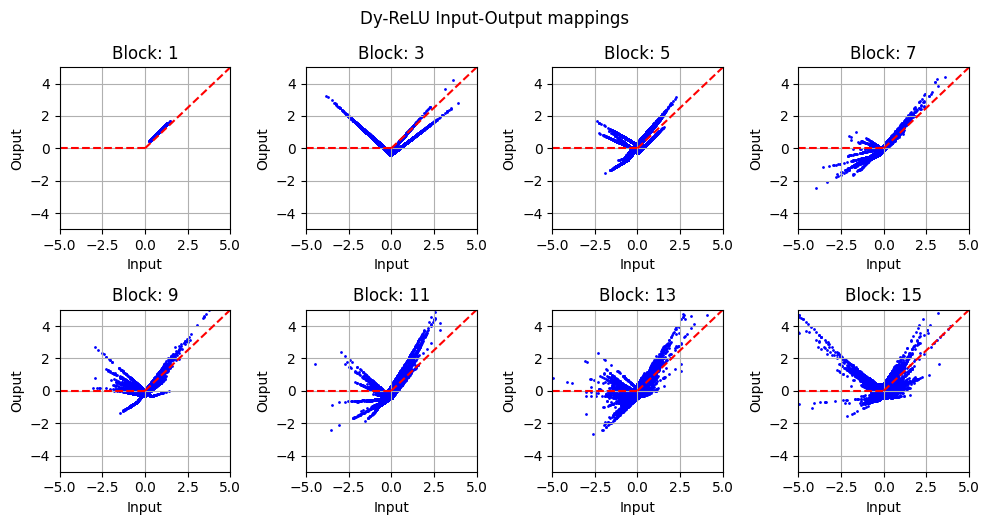}}
\caption{The figure shows the input to output mapping of Dy-ReLU for blocks at different depths in a well-trained DyMN-M. To generate the plots we use 10,000 randomly selected samples from the AudioSet evaluation set. Block 1 corresponds to the first block and Block 15 corresponds to the final block in Dy-MN. The dashed red line depicts the conventional ReLU function.}% caption command
\label{fig:dyrelu_mapping}% label
\end{figure*}

\subsection{Dy-Conv}
\label{subsec:inspecting_dy_conv}

In addition to \textit{context shuffle}, we probe for input-dependency of Dy-Conv by randomly shuffling the $K$ entries in the attention weight vectors (\textit{attention shuffle}). Both methods lead to a moderate performance decrease, showing that Dy-Conv is the least input-dependent operation of the three dynamic components in our setup. We further investigate the diversity of the $K$ dynamic kernels learned by constructing the kernel $W$ using \textit{uniform attention} weights ($W=\sum_k^K W_k/K$) and selecting the kernel $W$ based on the \textit{max attention} weight ($W=W_{\mathrm{argmax}_k(\alpha_k)}$). Since the performance in case of \textit{uniform attention} drops only by 2.5 points in mAP, we conclude that the learned kernels are similar to each other. However, when we conduct the same experiment for a model using Dy-Conv as the only dynamic component (such as the setting \mbox{\textit{-- CA, Dy-ReLU}} in Table~\ref{tab:abl_dyn_comp}), we find that the performance when using \textit{context shuffle} or \textit{uniform attention} drops to a value close to random guessing. This indicates that Dy-ReLU and CA already perform much of the capabilities of Dy-Conv. This is also aligned with the finding that Dy-Conv leads to the smallest performance improvement among the three dynamic methods, as shown in Table~\ref{tab:abl_dyn_comp}.

\subsection{Dy-ReLU}
\label{subsec:inspecting_dy_relu}

In case of Dy-ReLU, \textit{context shuffle} leads to a substantial performance drop of 15.3 points mAP. However, if we shuffle the predicted coefficients randomly across channels (\textit{channel shuffle}), the performance drop is much more severe (44.3 points mAP). These results show that Dy-ReLU learns a prior for the channel coefficients shared across the samples, aligned with the results of CA. 

Fig.~\ref{fig:dyrelu_mapping} provides additional insights into the behaviour of Dy-ReLU. It shows the Dy-ReLU input to output mapping of several blocks at different depth in a well trained \mbox{DyMN-M}. The input to output mappings are collected from 10,000 randomly drawn samples from the AudioSet evaluation set. The dashed red line indicates the conventional static ReLU function. We make sure that patterns described in the following hold across several trained DyMNs and not only for the model used to create Fig.~\ref{fig:dyrelu_mapping}. An interesting pattern can be detected in Block 1. The input values are exclusively within a small range of positive values and the Dy-ReLU approximates an identity function. In general, Dy-ReLUs in early blocks tend to learn to map points at lines with specific slopes. For instance, the shape of the mappings in Block 3 resembles the absolute value function. In contrast, Dy-ReLUs in later blocks, such as Blocks 13 and 15, show a highly dynamic behaviour, mapping specific input values to a wide range of different output values. Different to the conventional ReLU, Dy-ReLU maps a lots of negative input values to positive activations.

\section {Conclusion}
\label{sec:conlusion}

In this work, we proposed dynamic convolutional neural networks as efficient pre-trained audio models. We integrated dynamic convolutions, dynamic ReLU, and Coordinate Attention into efficient inverted residual blocks and share the computation of a global context for dynamic parameterization across all dynamic modules in a block. The resulting models, named DyMNs, are pre-trained on AudioSet at three different complexity levels using Transformer-to-CNN Knowledge Distillation. DyMNs show a beneficial performance--complexity trade-off compared to their non-dynamic counterparts and other Transformers and CNNs. Specifically, Dy-MN-L achieves a pre-training performance of 49.0 mAP on AudioSet, outperforming current popular Audio Spectrogram Transformers. Experiments on downstream tasks indicate that the proposed DyMNs outperform other CNNs by a large margin and are highly competitive compared to Audio Spectrogram Transformers while being much more computationally efficient. Furthermore, we show that DyMNs are suitable for simple task-specific fine-tuning by sharing the same fine-tuning pipeline across all downstream tasks. In short, DyMNs are efficient, high-performing, easy-to-fine-tune audio models that can have a large impact on the audio community, especially in the context of resource-critical applications.

\section{ACKNOWLEDGMENT}

The computational results presented were achieved in part using the Linz Institute of Technology (LIT) AI Lab Cluster. The LIT AI Lab is supported by the Federal State of Upper Austria. Gerhard Widmer's work is supported by the European Research Council (ERC) under the European Union's Horizon 2020 research and innovation programme, grant agreement No 101019375 (Whither Music?).

\bibliographystyle{IEEEtran}
\bibliography{refs}

\newpage

\vfill

\end{document}